\documentclass[12pt,dvips]{article}

\usepackage{rotating}
\usepackage{epsfig}
\usepackage{color}
\usepackage{cite}

\newcommand{\real}{\Re {\rm e}}

\newcommand{\lsim}{\raisebox{-0.13cm}{~\shortstack{$<$ \\[-0.07cm] $\sim$}}~}
\newcommand{\gsim}{\raisebox{-0.13cm}{~\shortstack{$>$ \\[-0.07cm] $\sim$}}~}

\begin{document}

\def\thefootnote{\fnsymbol{footnote}}

{\small
\begin{flushright}
CNU-HEP-15-07,
FERMILAB-PUB-15-508-T,
EFI-15-36 \\
MAN/HEP/2015/19,
KCL-PH-TH/2015-49, LCTS/2015-37, CERN-PH-TH/2015-252
\end{flushright} }

\medskip

\begin{center}
{\bf 
\hspace{-1cm}{\Large CP Violation in Heavy MSSM Higgs Scenarios} 
}
\end{center}

\smallskip

\begin{center}{\large
M.~Carena$^{\,a,b,c}$, 
J.~Ellis$^{d,e}$, 
J.~S.~Lee$^{f}$,
A.~Pilaftsis$^{e,g}$, 
and C.~E.~M.~Wagner$^{b,c,h}$}
\end{center}

\begin{center}
{\em $^a$Fermi National Accelerator Laboratory, P.O. Box 500, Batavia IL 60510, U.S.A.}\\[0.2cm]
{\em $^b$Enrico Fermi Institute, University of Chicago, Chicago, IL 60637, U.S.A.}\\[0.2cm]
{\em $^c$Kavli Institute for Cosmological Physics, University of Chicago, Chicago, IL 60637, U.S.A.}\\[0.2cm]
{\em $^d$Theoretical Particle Physics and Cosmology Group, Department of 
  Physics, King's~College~London, London WC2R 2LS, United Kingdom}\\[0.2cm]
{\em $^e$Theory Division, CERN, CH-1211 Geneva 23, Switzerland}\\[0.2cm]
{\em $^f$Department of Physics, Chonnam National University, \\
300 Yongbong-dong, Buk-gu, Gwangju, 500-757, Republic of Korea}\\[0.2cm]
{\em $^g$Consortium for Fundamental Physics, School of Physics and Astronomy}\\
{\em University of Manchester, Manchester M13 9PL, United Kingdom}\\[0.2cm]
{\em $^h$HEP Division, Argonne National Laboratory,
9700 Cass Ave., Argonne, IL 60439, USA}\\
\end{center}

\bigskip

\centerline{\bf ABSTRACT} {\small \medskip\noindent We introduce and
  explore new heavy Higgs scenarios in the Minimal Supersymmetric
  Standard Model (MSSM) with explicit CP violation, which have
  important phenomenological implications that may be testable at the
  LHC.  For soft supersymmetry-breaking scales M$_S$ above a few TeV
  and a charged Higgs boson mass~$M_{H^+}$ above a few hundred GeV,
  new physics effects including those from explicit CP violation
  decouple from the light Higgs boson sector.  However, such effects
  can significantly alter the phenomenology of the heavy Higgs bosons
  while still being consistent with constraints from low-energy
  observables, for instance electric dipole moments.  To consider
  scenarios with a charged Higgs boson much heavier than the Standard
  Model~(SM) particles but much lighter than the supersymmetric
  particles, we revisit previous calculations of the MSSM Higgs
  sector. We compute the Higgs boson masses in the presence of CP
  violating phases, implementing improved matching and
  renormalization-group~(RG) effects, as well as two-loop RG effects
  from the effective two-Higgs Doublet Model (2HDM) scale $M_{H^\pm}$
  to the scale $M_S$.  We illustrate the possibility of non-decoupling
  CP-violating effects in the heavy Higgs sector using new benchmark
  scenarios named \textcolor{red}{CP}\textcolor{green}{X}{\it
    4}\textcolor{blue}{LHC}.}

\vspace{0.2in}
\newpage

\section{Introduction}
\label{sec:intro}

Supersymmetry (SUSY)  remains one of the  best-motivated extensions of
the  Standard Model  (SM), despite  the current  lack of  evidence for
supersymmetric  partner  particles at  the  LHC.   In particular,  the
discovery  of a  light Higgs  boson in Run  I of  the LHC,  is in
agreement  with the  predictions from  SUSY.  Supersymmetric  theories
provide   a  viable   mechanism  for   stabilizing  the   electroweak
vacuum~\cite{ER} and  require a restricted  range for the mass  of the
lightest   Higgs   boson~\cite{ERZ,HH,OYT}   that   contains   the   measured
value~\cite{MH}.  Moreover, minimal low-energy SUSY models with masses
of  the additional  Higgs bosons  and supersymmetric  particles larger
than the  weak scale lead  to values  of the lightest  Higgs couplings
that are close to the SM ones~\cite{EHOW}, as suggested by current LHC
experiments~\cite{muH}.

On the other hand,  the non-discovery of SUSY in Run I  of the LHC has
disproved  benchmark  scenarios  proposed  previously~\cite{CPX},  and
motivates the  consideration of new  benchmarks that can be  tested in
future runs of the LHC.  Specifically, it is plausible to consider the
case that the common soft SUSY-breaking scale $M_S$ is $ \gsim 2$~TeV,
whereas  the mass  scale  $M_H$ of  the heavy  MSSM  Higgs bosons,  as
determined  by  the  charged  Higgs boson  mass  $M_{H^+}$,  could  be
somewhat lower, in  the few to several hundred GeV  range. In relation
to  this, we  recall that  future runs  of the  LHC at  13/14 TeV  are
expected   to   be  sensitive   to   squarks   and  gluinos   weighing
$\lsim  3$~TeV, and  heavy MSSM  Higgs bosons  weighing $\lsim  2$~TeV
depending on the value of $\tan \beta$.  Accordingly, in this paper we
introduce  and  explore  MSSM  Higgs boson  benchmark  scenarios  with
$100~{\rm GeV} \ll M_H \ll M_S \gsim 2$~TeV.

Our principal interest in new heavy Higgs boson benchmark scenarios is
the possible manifestation of observable CP violation in the Higgs
sector of the MSSM. It is well known~\cite{Pilaftsis:1998pe,
  Pilaftsis:1998dd, Pilaftsis:1999qt, Demir:1999hj, Choi:2000wz,
  Carena:2000yi, Ibrahim:2000qj, Carena:2001fw,
  Ibrahim:2002zk} that such a possibility arises in the
MSSM Higgs potential beyond the tree-level approximation,
predominantly from CP phases in the soft SUSY-breaking trilinear
couplings of stops and sbottoms, but also from CP phases in the
gaugino masses.  However, experimental upper limits on electric dipole
moments (EDMs) severely constrain the size of such CP-violating
parameters as predicted in the MSSM at one-, two- and higher
loops~\cite{ELPedm1}.  In particular, in~the absence of cancellations
between these different contributions~\cite{Ibrahim:1998je,
  Brhlik:1998zn} as occur along specific directions in the space of
CP-odd phases~\cite{ELPedm2,Yamanaka}, the EDM constraints effectively
preclude the observation of CP-violating effects in the couplings of
the Higgs boson discovered at the LHC~\cite{Li:2015yla}. However, the
observation of CP violation effects elsewhere, notably in the heavy
MSSM Higgs
sector~\cite{Pilaftsis:1998dd,Pilaftsis:1999qt,Carena:2000yi} or
$B$-meson decays~\cite{Ellis:2007kb,Arbey:2014msa} is not excluded.
These CP-violating effects have often been studied in the framework of
the CPX scenarios proposed previously~\cite{CPX}, but in light of the
LHC Run-I limits on supersymmetric particle masses and the observed
Higgs boson properties, the CPX benchmarks should be revisited.

With the above motivations in mind, in this paper we present new precision
calculations  of  the  MSSM  Higgs  spectrum in  the  presence  of  CP
violation, which  are suitable for  scenarios in which the  SUSY scale
$M_S$  is (far)  beyond the  TeV region.   To this  end, we  solve the
two-loop RGEs of the two-Higgs-doublet  model (2HDM) in the range $M_S
> Q >  M_H$, as well as the two-loop  SM RGEs in the range  $M_H > Q >
m_t^{\rm pole}$, implementing full one-loop matching conditions at the
relevant  thresholds  $M_S,  M_H$   and  $m_t^{\rm  pole}$.   All  the
improvements considered here  are being implemented in a  new version of
the   public   code   {\tt  CPsuperH}~\cite{Lee:2003nta,   Lee:2007gn,
  Lee:2012wa}, namely {\tt  CPsuperH3.0}~\footnote{For another tool to
  calculate      CP-violating      effects      in      the      MSSM,
  see~\cite{FeynHiggs,Hahn:2013ria}.}.  The  full description with all
the detailed  information about  {\tt  CPsuperH3.0} will be presented  in a
future publication.

Section~2 of this paper reviews  the conventions and notations of {\tt
  CPsuperH}  that we  use  for our  analysis, as  well  as some  basic
formulae for the Higgs  boson self-energies.  Section~3 specifies
the  matching   conditions  and  the   RG  running  effects   that  we
incorporate.   In section~4  we present some numerical results for the Higgs
spectra. In
section~5 we introduce
our new  CP-violating  benchmark
scenarios           (\textcolor{red}{CP}\textcolor{green}{X}{\it
  4}\textcolor{blue}{LHC})  for  the  MSSM  heavy  Higgs  sector,  and
 present          the          results          for the
\textcolor{red}{CP}\textcolor{green}{X}{\it
  4}\textcolor{blue}{LHC} benchmarks.   Our conclusions  are summarized
in Section~6.  The main text of the paper is accompanied by Appendices
containing  detailed formulae:  Appendix~A  contains  the relevant  SM
RGEs, Appendix~B contains the one--loop 2HDM RGEs, Appendix~C contains
the  two-loop  2HDM  RGEs,  and Appendix~D  summarizes  the  threshold
corrections to quartic couplings at the scale $M_S$.

\section{The CP-Violating MSSM Higgs Sector}

In this section we review the computation of the Higgs boson
self-energies and pole masses and record the basic expressions used in
{\tt CPsuperH3.0}, that underlie our present analysis.  We follow the
conventions and notations of {\tt
  CPsuperH}~\cite{Lee:2003nta,Lee:2007gn,Lee:2012wa}, unless stated
otherwise explicitly.

\subsection{The Two-Higgs-Doublet Model (2HDM)}

The tree-level 2HDM Higgs potential can be written as~\cite{Pilaftsis:1999qt}:
\begin{eqnarray}
  \label{LV}
{\cal L}_V &=& \mu^2_1 (\Phi_1^\dagger\Phi_1)\, +\,
\mu^2_2 (\Phi_2^\dagger\Phi_2)\, +\, m^2_{12} (\Phi_1^\dagger \Phi_2)\,
+\, m^{*2}_{12} (\Phi_2^\dagger \Phi_1)\, +\,
\lambda_1 (\Phi_1^\dagger \Phi_1)^2\, +\,
\lambda_2 (\Phi_2^\dagger \Phi_2)^2\,\nonumber\\
&& +\, \lambda_3 (\Phi_1^\dagger \Phi_1)(\Phi_2^\dagger \Phi_2)\, +\,
\lambda_4 (\Phi_1^\dagger \Phi_2)(\Phi_2^\dagger \Phi_1)\, +\,
\lambda_5 (\Phi_1^\dagger \Phi_2)^2\, +\,
\lambda^*_5 (\Phi_2^\dagger \Phi_1)^2\, \\
&&+\, \lambda_6 (\Phi_1^\dagger \Phi_1)(\Phi_1^\dagger \Phi_2)\, +
\lambda^*_6 (\Phi_1^\dagger \Phi_1)(\Phi_2^\dagger \Phi_1)\,
+\, \lambda_7 (\Phi_2^\dagger \Phi_2)(\Phi_1^\dagger \Phi_2)\, +
\lambda^*_7 (\Phi_2^\dagger \Phi_2)(\Phi_2^\dagger \Phi_1)\, .
\nonumber
\end{eqnarray}
The relations between these and the conventional MSSM parameters are
\begin{eqnarray}
  \label{LVpar}
\mu^2_1 &=& -m^2_1 - |\mu|^2\, ,\qquad \mu^2_2\ =\ -m^2_2 - |\mu|^2\, ,
                                                            \qquad
\lambda_1\ =\ \lambda_2\ =\ -\, \frac{1}{8}\, (g^2 + g'^2)\, ,\nonumber\\
\lambda_3 &=& -\frac{1}{4}\, (g^2 -g'^2)\, ,\qquad
\lambda_4\ =\ \frac{1}{2}\, g^2\, , \qquad \lambda_5\ =\ \lambda_6\
=\ \lambda_7\ =\ 0\, .
\end{eqnarray}
The doublet Higgs fields may be decomposed as follows:
\begin{equation}
  \label{Phi12}
\Phi_1\ =\ \left( \begin{array}{c}
\phi^+_1 \\ \frac{1}{\sqrt{2}}\, ( v_1\, +\, \phi_1\, +\, ia_1)
\end{array} \right)\, ,\qquad
\Phi_2\ =\ e^{i\xi}\, \left( \begin{array}{c}
\phi^+_2 \\  \frac{1}{\sqrt{2}}\, ( v_2 \, +\, \phi_2\, +\, ia_2 )
 \end{array} \right)\, ,
\end{equation}
where the charged and neutral Goldstone bosons, $G^\pm$ and $G^0$, are
determined through the relations:
\begin{equation}
  \label{rot}
\left(\! \begin{array}{c} G^+ \\ H^+ \end{array}\!\right)\ = \
\left(\! \begin{array}{cc} c_\beta & s_\beta \\
-\,s_\beta & c_\beta \end{array}\!\right)\,
\left(\! \begin{array}{c} \phi^+_1 \\ \phi^+_2 \end{array}\!\right)\,,\qquad
\left(\! \begin{array}{c} G^0 \\ a \end{array}\!\right)\ = \
\left(\! \begin{array}{cc} c_\beta & s_\beta \\
-\,s_\beta & c_\beta \end{array}\!\right)\, \left(\! \begin{array}{c}
a_1 \\ a_2 \end{array}\!\right) \,,
\end{equation}
with $s_\beta \equiv \sin\beta$, $c_\beta \equiv \cos\beta$ and $\tan\beta = v_2/v_1$.

To  make contact  with the  notations used  in~\cite{Haber:1993an}, we
make     the    following     identifications:    $H_u=\Phi_2$     and
$H_d=\widetilde\Phi_1=i\tau_2\Phi_1^*=(\phi_1^{0*},-\phi_1^-)^T$.
Moreover, the  kinematic parameters as  defined in~\cite{Haber:1993an}
are related to ours as follows:
\begin{eqnarray}
\label{eq:match.hh.1}
&&
m_{11}^2 \to -\mu_1^2\,, \ \ \
m_{22}^2 \to -\mu_2^2\,, \ \ \
m_{12}^2 \to +m_{12}^2\,, 
\nonumber \\[2mm]
&&
\lambda_1 \to -2\lambda_1\,, \ \ \
\lambda_2 \to -2\lambda_2\,, \ \ \
\lambda_3 \to -\lambda_3\,, \ \ \
\lambda_4 \to -\lambda_4\,, 
\nonumber \\[2mm]
&&
\lambda_5 \to -2\lambda_5\,, \ \ \
\lambda_6 \to -\lambda_6\,, \ \ \
\lambda_7 \to -\lambda_7\,,
\nonumber \\[2mm]
&&
g_2 \to g \,, \ \ \
g_1 \to g^\prime \,. \ \ \
\cdots
\end{eqnarray}
The one-loop 2HDM RGEs are given in Appendix~B~\footnote{We note 
that the RGE running parameter used in Ref.~\cite{Haber:1993an}
is related to ours by $t \to 2t$.}, and the two-loop 2HDM RGEs
are given in Refs.~\cite{Dev:2014yca},\cite{Lee:2015uza} and Appendix C.

\subsection{Charged Higgs Bosons}

In the $\{\phi_1^\pm,\phi_2^\pm\}$ basis, the RG-improved charged
Higgs-boson self-energy matrix can be found in Eq.~(2.6) of
Ref.~\cite{Carena:2001fw}:
\begin{equation}
   \label{Pihatplus}
\left(\widehat{\Pi}^\pm\right)_{ij} (s)\ =\ -\left(\overline{{\cal M}}_\pm^2\right)_{ij}
+(\xi^+_i\xi^-_j)^{-1} \left(\Delta\Pi^\pm\right)^{\tilde{f}}_{ij}(s)
+\left(\widetilde{\Pi}^\pm\right)^{f}_{ij}(s)\; .
\end{equation}
The first term,  $\overline{{\cal M}}_\pm^2$, is the two-loop Born-improved squared-mass matrix,
\begin{equation}
\overline{{\cal M}}_\pm^2\ =\ \left(\frac{1}{2}\bar\lambda_4 v_1 v_2
+\real{\overline{m}_{12}^2}\right)\left(\begin{array}{cc}
\tan\beta & -1 \\ -1 & \cot\beta
\end{array} \right) \, ,
\end{equation}
expressed in terms of relevant parameters such as the real part of the
soft bilinear Higgs mixing, $\real{\overline{m}_{12}^2}$, and the
quartic coupling $\lambda_4$. The bar on these parameters indicates
the sum of the tree-level and of the one- and two-loop leading
logarithmic contributions.  When solving the 2HDM RGEs,
$\bar\lambda_4$ is to be estimated at the scale~$M_H$ where the heavy
Higgs bosons decouple, and $\real{\overline{m}_{12}^2}$ is fixed when
the charged-Higgs-boson pole mass is given as an input, as shown
below.

The second term in~(\ref{Pihatplus}) describes the threshold effects
of the sfermions (top and bottom squarks) and is the product of two
quantities: (i) the anomalous dimension factors $\xi_i$
\begin{equation}
\xi_i=\exp\left[-\int_{\ln M_H}^{\ln M_S}\gamma_i(t)\,{\rm d}t\right] \, ,
\label{eq:xii}
\end{equation}
defined in terms of the anomalous-dimensions of the external
Higgs fields $\gamma_i \equiv {\rm d}\ln\Phi_i/{\rm d}t$ (in this case
the charged Higgs fields), and (ii) the scale-invariant one-loop
threshold contribution from the top and bottom squarks
\begin{equation}
   \label{DeltaPipm}
\left(\Delta\Pi^\pm\right)^{\tilde{f}}\ =\ \left(\frac{1}{2}\lambda_4^{(1)} v_1 v_2
+\real{m_{12}^{2(1)}}\right)\left(\begin{array}{cc}
\tan\beta & -1 \\ -1 & \cot\beta
\end{array} \right)
\: +\: \left(\widetilde{\Pi}^\pm\right)^{\tilde{f}}\; .
\end{equation}
In the above, the SUSY-breaking scale $M_S$ is used to decouple the
heavy sfermions. Moreover, the superscript \lq\lq(1)" in
$\lambda_4^{(1)}$ and $\real{m_{12}^{2(1)}}$ indicates that these
quantities contain the one-loop leading logarithmic contributions and
they can be obtained from Eqs.~(3.6) and (3.7) of~\cite{Carena:2000yi}
by choosing $Q=M_S$.

We note  that the  vacuum expectation values  (VEVs) $v_{1,2}$  of the
Higgs doublets $\Phi_{1,2}$, and hence  $\tan \beta$, evolve with the
wave-function renormalization  factors $\xi_{1,2}$  of the
corresponding neutral Higgs bosons:
\begin{equation}
v_i(M_S)\ =\ v_i(M_H)/\xi_i\; ,\qquad
\tan\beta(M_S)\ =\ \tan\beta(M_H)\frac{\xi_1}{\xi_2} \; ,
\end{equation}
where $\tan\beta (M_H)$ is the input value of $\tan\beta$, i.e.~at the
scale $Q=M_H$.  Consequently,  the SM VEV $v$ is related  to the Higgs
VEVs $v_{1,2}$ through:
\begin{equation}
v_1(M_H)\ =\ c_\beta(M_H)\, v(M_H)\;, \qquad v_2(M_H)\ =\ s_\beta(M_H)\, v(M_H)\; .
\end{equation}
 The SM VEV $v$ is fixed at the RG scale $Q = m_t$, by virtue of the
 relation: $v(M_H)\, =\, v(m_t)/\xi_{\rm SM}$, where
\begin{equation}
\xi_{\rm SM}\ =\ 
\exp\left[-\int_{\ln m_t}^{\ln M_H}\gamma(t)\,{\rm d}t\right]\; .
\end{equation}
Here $\gamma (t)$ is the anomalous dimension of the SM Higgs doublet,
which is given in~(\ref{gammaSM}) in the one-loop approximation.

Finally,   the   last   terms   on  the   RHSs   of~(\ref{DeltaPipm})
and~(\ref{Pihatplus}),                                           namely
$\left(\widetilde{\Pi}^\pm\right)^{\tilde{f}}$                      and
$\left(\widetilde{\Pi}^\pm\right)^f$,  can be expressed as
follows, 
\begin{equation}
  \label{RenSEchargedMSbar}
\left(\widetilde{\Pi}^\pm\right)^{\tilde{f}/f}\ =\
\left(\Pi^\pm\right)^{\tilde{f}/f}\: +\: \left(\begin{array}{cc} 
\frac{\left( T_{\phi_1}\right)^{\tilde{f}/f}}{v_1} & i\frac{\left(
    T_a\right)^{\tilde{f}/f}}{v} \\ 
-i\frac{\left( T_a\right)^{\tilde{f}/f}}{v} &
\frac{\left(T_{\phi_2}\right)^{\tilde{f}/f}}{v_2}  
\end{array}\right)\; ,
\end{equation}
with all quantities in the RHS of~(\ref{RenSEchargedMSbar}) computed
in the $\overline{\rm MS}$ scheme.  Explicit one-loop calculations yield
\begin{eqnarray}
\left(\Pi^\pm\right)^{\tilde{f}} &=& 
\Pi^{\pm\,(a)} + \Pi^{\pm\,(b)}\; ,\qquad
\left(T_{\phi_{1,2}}\right)^{\tilde{f}}\ =\ T_{\phi_{1,2}}^{(d)}\;,\qquad
\left(T_{a_{1,2}}\right)^{\tilde{f}}\ =\ T_{a_{1,2}}^{(d)}\;,\qquad
\nonumber \\[2mm]
\left(\Pi^\pm\right)^f &=& \Pi^{\pm\,(c)}\;, \qquad\qquad\qquad\hspace{-1mm}
\left(T_{\phi_{1,2}}\right)^f\ =\ T_{\phi_{1,2}}^{(e)} \; ,\qquad 
\left(T_{a_{1,2}}\right)^f\ =\ 0\;, \qquad
\end{eqnarray}
with $T_a=T_{a_2}/c_\beta=-T_{a_1}/s_\beta$ and where
$\Pi^{\pm\,(a)}$, $\Pi^{\pm\,,(b)}$, $\Pi^{\pm\,(c)}$,
$T_{\phi_{1,2}}^{(d)}$, $T_{a_{1,2}}^{(d)}$, and
$T_{\phi_{1,2}}^{(e)}$ are given by Eqs.~(B.12), (B.13), (B.15), and
(B.16) in~\cite{Carena:2001fw}.  The sfermionic contributions should
be calculated at the scale $M_S$, whereas the fermionic contributions
are evaluated at $M_H$.

In the $\{G^\pm,H^\pm\}$  basis, the inverse-propagator  matrix of the
charged Higgs bosons is given by
\begin{equation}
\label{InverseProp}
\widehat\Delta_\pm^{-1}(s) = s\,{\bf 1}_{2\times 2} +
\left(\begin{array}{cc}
c_\beta & s_\beta \\ -s_\beta & c_\beta 
\end{array} \right)\, 
\widehat\Pi^\pm(s) \,
\left(\begin{array}{cc}
c_\beta & -s_\beta \\ s_\beta & c_\beta 
\end{array} \right) \, ,
\end{equation}
where we have defined
\begin{equation}
   \label{Pihatplus1}
\left(\widehat{\Pi}^\pm\right)_{ij} (s)\ =\
 -\left(\overline{{\cal M}}_\pm^2\right)_{ij} 
+\left(\Delta\widehat{\Pi}^\pm\right)_{ij}(s) \, .
\end{equation}
In (\ref{InverseProp}), the $\{22\}$ matrix element of the second term is given by
\begin{eqnarray}
&&
 \left(\widehat\Pi^\pm\right)_{11}\,s_\beta^2
-\left[\left(\widehat\Pi^\pm\right)_{12}
+\left(\widehat\Pi^\pm\right)_{21}\right]\,s_\beta c_\beta
+\left(\widehat\Pi^\pm\right)_{22}\,c_\beta^2
\nonumber \\[2mm]
&&\hspace{7cm}=\ 
-\left(\frac{1}{2}\bar\lambda_4 v^2 +
\frac{\real\overline{m}_{12}^2}{c_\beta s_\beta} \right)
+\Delta\widehat\Pi_{H^+H^-} \, ,\qquad
\end{eqnarray}
with
\begin{equation}
\Delta\widehat\Pi_{H^+H^-} \ \equiv\
\left(\Delta\widehat\Pi^\pm\right)_{11}\,s_\beta^2
-\left[\left(\Delta\widehat\Pi^\pm\right)_{12}
+\left(\Delta\widehat\Pi^\pm\right)_{21}\right]\,s_\beta c_\beta
+\left(\Delta\widehat\Pi^\pm\right)_{22}\,c_\beta^2 \, .
\end{equation}
This yields the  pole mass condition
\begin{eqnarray}
&&
\real\left(\widehat\Delta_\pm^{-1}\right)_{22}(s=M_{H^\pm}^2)
\nonumber \\ &&
\hspace{2cm} =\ M_{H^\pm}^2\: -\: \left(\frac{1}{2}\bar\lambda_4 v^2\: +\:
\frac{\real\overline{m}_{12}^2}{c_\beta s_\beta} \right)\:
+\: \real\Delta\widehat\Pi_{H^+H^-}(s=M_{H^\pm}^2)\ =\ 0\,, \quad
\end{eqnarray}
which may be used to eliminate $\real\overline{m}_{12}^2$ in favor of the
charged-Higgs boson pole mass $M_{H^\pm}^2$.

\subsection{Neutral Higgs Bosons}

In the $\{\phi_1,\phi_2,a_1,a_2\}$ basis, Eq.~(2.14) of \cite{Carena:2001fw} takes the form
\begin{equation}
\widehat\Pi^N(s)\ =\ \left(\begin{array}{cc}
\widehat\Pi^S(s)  &
\widehat\Pi^{SP}(s)  \\
\left(\widehat\Pi^{SP}(s)\right)^T  &
\widehat\Pi^P(s)  
\end{array}\right) \, ,
\end{equation}
with
\begin{eqnarray}
\left(\widehat{\Pi}^S\right)_{ij} (s)
&=& -\left(\overline{{\cal M}}_S^2\right)_{ij}
+(\xi_i\xi_j)^{-1} \left(\Delta\Pi^S\right)^{\tilde{f}}_{ij}(s)
+\left(\widetilde{\Pi}^S\right)^{f}_{ij}(s) \, ,
\nonumber \\[2mm]
\left(\widehat{\Pi}^P\right)_{ij} (s)
&=& -\left(\overline{{\cal M}}_P^2\right)_{ij}
+(\xi_i\xi_j)^{-1} \left(\Delta\Pi^P\right)^{\tilde{f}}_{ij}(s)
+\left(\widetilde{\Pi}^P\right)^{f}_{ij}(s) \, ,
\nonumber \\[2mm]
\left(\widehat{\Pi}^{SP}\right)_{ij} (s) &=&
(\xi_i\xi_j)^{-1} \left(\widetilde\Pi^{SP}\right)^{\tilde{f}}_{ij}(s)
+\left(\widetilde{\Pi}^{SP}\right)^{f}_{ij}(s) \, .
\end{eqnarray}
where, in analogy with Eq.~(\ref{eq:xii}),  $\xi_i$ are the
corresponding anomalous dimension factors of the neutral Higgs
fields. 

The quantities $\overline{{\cal M}}_S^2$ and $\overline{{\cal M}}_P^2$ appearing here may
be written in the forms
\begin{eqnarray}
\overline{{\cal M}}_S^2 &= &
\real{\overline{m}_{12}^2}\,\left(\begin{array}{cc}
\tan\beta & -1 \\ -1 & \cot\beta
\end{array} \right) 
\: -\: v^2\,\left(\begin{array}{cc}
2\bar\lambda_1c^2_\beta & \bar\lambda_{34}c_\beta s_\beta \\ 
\bar\lambda_{34} c_\beta s_\beta & 2\bar\lambda_2 s^2_\beta 
\end{array} \right) \, ,
\nonumber \\[2mm]
\overline{{\cal M}}_P^2 &= &
\real{\overline{m}_{12}^2}\,\left(\begin{array}{cc}
\tan\beta & -1 \\ -1 & \cot\beta
\end{array} \right) \, ,
\end{eqnarray}
where $\bar\lambda_1$, $\bar\lambda_2$, and
$\bar\lambda_{34}=\bar\lambda_3+\bar\lambda_4$  are
to be evaluated by solving the 2HDM RGEs at the scale~$M_H$.

The quantities $\Delta\Pi^S$ and $\Delta\Pi^P$ may be written as
\begin{eqnarray}
\left(\Delta\Pi^S\right)^{\tilde{f}} &= &
\real{{m}_{12}^{2(1)}}\,\left(\begin{array}{cc}
t_\beta & -1 \\ -1 & 1/t_\beta
\end{array} \right) 
\: -\: v^2\,\left(\begin{array}{cc}
2\lambda_1^{(1)}c^2_\beta & \lambda_{34}^{(1)}c_\beta s_\beta \\ 
\lambda_{34}^{(1)}c_\beta s_\beta & 2\lambda_2^{(1)}s^2_\beta 
\end{array} \right) 
\: +\: \left(\widetilde{\Pi}^S\right)^{\tilde{f}} \, ,
\nonumber \\[2mm]
\left(\Delta\Pi^P\right)^{\tilde{f}} &= &
\real{m_{12}^{2(1)}}\,\left(\begin{array}{cc}
t_\beta & -1 \\ -1 & 1/t_\beta
\end{array} \right)
\: +\: \left(\widetilde{\Pi}^P\right)^{\tilde{f}} \, ,
\end{eqnarray}
where $\lambda_{1,2,34}^{(1)}$ and $\real{m_{12}^{2(1)}}$ can be obtained 
from Eqs.~(3.3), (3.4), (3.5), (3.6) and (3.7) of~\cite{Carena:2000yi} by choosing $Q=M_S$.

The   quantities   $\widetilde{\Pi}^{S,P,SP}$   are   given   in   the
$\overline{\rm MS}$ scheme by Eq.~(2.11) of \cite{Carena:2001fw}:
\begin{eqnarray}
\widetilde{\Pi}^S &=& \Pi^S + \left(\begin{array}{cc}
\frac{T_{\phi_1}}{v_1} & 0 \\
0 & \frac{T_{\phi_2}}{v_2} 
\end{array}\right) \, ,
\nonumber \\[2mm]
\widetilde{\Pi}^{SP} &=& \Pi^{SP} + \frac{T_a}{v}\left(\begin{array}{cc}
0 & +1 \\ -1 & 0 \end{array}\right) \, ,
\nonumber \\[2mm]
\widetilde{\Pi}^P &=& \Pi^P + \left(\begin{array}{cc}
\frac{T_{\phi_1}}{v_1} & 0 \\
0 & \frac{T_{\phi_2}}{v_2} 
\end{array}\right) \, .
\end{eqnarray}
Here,  $\left(\Pi^S\right)^{\tilde f}$ and  $\left(\Pi^S\right)^f$ are
given by
\begin{equation}
\left(\Pi^S\right)^{\tilde f} = \Pi^{S,(a)} + \Pi^{S,(b)}\,,  \qquad
\left(\Pi^S\right)^f = \Pi^{S,(c)}\; , 
\end{equation}
which    are   specified    in   Eqs.~(B.5),    (B.6),    and   (B.14)
of~\cite{Carena:2001fw}.  In addition, $\left(\Pi^P\right)^{\tilde f}$
and $\left(\Pi^P\right)^f$ are given by
\begin{equation}
\nonumber \\[2mm]
\left(\Pi^P\right)^{\tilde f} = \Pi^{P,(a)} + \Pi^{P,(b)}\,,  \qquad
\left(\Pi^P\right)^f = \Pi^{P,(c)} \; ,
\end{equation}
which can be obtained from Eqs.~(B.5), (B.6) by replacing  $\phi_i \to a_i$
and from Eq.~(B.14) of \cite{Carena:2001fw}. Moreover, the
CP-violating self-energies $\left(\Pi^{SP}\right)^{\tilde f}$ and
 $\left(\Pi^{SP}\right)^f$ may be expressed as
\begin{equation}
\left(\Pi^{SP}\right)^{\tilde f} = \Pi^{{SP},(a)} \,,  \qquad
\left(\Pi^{SP}\right)^f = 0 \;. 
\end{equation}
The non-zero  self-energy $\Pi^{{SP},(a)}$  is given by  Eq.~(B.11) of
\cite{Carena:2001fw}.

Finally, the inverse propagator matrix  of the neutral Higgs bosons in
the $\{\phi_1,\phi_2,a,G^0\}$ basis is given by
\begin{equation}
\widehat\Delta_N^{-1}(s) = s\,{\bf 1}_{4\times 4} +
\left(\begin{array}{cccc}
1 & 0 & 0 & 0 \\
0 & 1 & 0 & 0 \\
0 & 0 & -s_\beta & c_\beta \\
0 & 0 &  c_\beta & s_\beta \\
\end{array} \right)\, 
\widehat\Pi^N(s) \,
\left(\begin{array}{cccc}
1 & 0 & 0 & 0 \\
0 & 1 & 0 & 0 \\
0 & 0 & -s_\beta & c_\beta \\
0 & 0 &  c_\beta & s_\beta \\
\end{array} \right)\, ,
\end{equation}
and  the   physical  masses  can   be  obtained  from   the  pole-mass
conditions. We should reiterate here that the parameter $\tan\beta$ is
defined at $s = 0$. In this kinematic limit, the Goldstone boson $G^0$
decouples from the $4\times 4$ propagator matrix, independently of the
presence      of      explicit      CP      violation      in      the
theory~\cite{Pilaftsis:1998pe},  as  a  consequence of  the  Goldstone
theorem.

\section{Matching Conditions and RG Running Effects}

Here we detail the $\overline{\rm MS}$ renormalization group approach
that we follow for the computation of the masses and mixings of the
neutral and charged Higgs bosons in the CP-violation case. In
particular, we state explicitly our matching conditions at the
relevant threshold scales. Given these matching conditions, we compute
the RG running effects to the relevant gauge, Yukawa and quartic
couplings between the different threshold scales.

To start with, we define the SUSY-breaking scale $M_S$ by
\begin{equation}
   \label{defineMS}
 M_S^2\ \equiv\ {\rm max}\left(M^2_{\widetilde Q_3}+m_t^2, M^2_{\widetilde U_3}+m_t^2,
\; M^2_{\widetilde D_3}+m_b^2, M^2_{\widetilde L_3}+m_\tau^2,
\; M^2_{\widetilde E_3}+m_\tau^2\right) \, ,
\end{equation}
which acts as the SUSY threshold scale. For the purposes of this study, we
ignore possible hierarchies between the third-generation sfermions,
by assuming they are small as compared to the other two hierarchical
scales: (i) the heavy Higgs-sector scale $M_H\equiv M_{H^+}$; (ii) the
top-quark mass $m_t$.

The matching conditions for the quartic and Yukawa couplings at the threshold
$M_S$ are as follows:
\begin{eqnarray}
&&
\bar\lambda_1 = \bar\lambda_2\ =\ -\, \frac{1}{8}\, (g^2 + g'^2)\,, \ \ \
\bar\lambda_3 = -\frac{1}{4}\, (g^2 -g'^2)\,,\ \ \
\bar\lambda_4 = \frac{1}{2}\, g^2 \,;
\nonumber \\[2mm] &&
h_t^{\rm MSSM}=\frac{h_t^{\rm 2HDM}}{1+\delta_t +\cot\beta \Delta_t} \, ,
\nonumber \\[2mm] &&
h_b^{\rm MSSM}=\frac{h_b^{\rm 2HDM}}{1+\delta_b +\tan\beta \Delta_b} \, ,
\nonumber \\[2mm] &&
h_\tau^{\rm MSSM}=\frac{h_\tau^{\rm 2HDM}}{1+\tan\beta \Delta_\tau} \, ,
\end{eqnarray}
where      $\Delta_f      =      \Delta      h_f/h_f^{\rm      MSSM}$,
$\delta_f  =   \delta  h_f/h_f^{\rm  MSSM}$,  and   $\Delta  h_f$  and
$\delta h_f$ are the supersymmetric threshold corrections to the third
generation  Yukawa couplings~\cite{Carena:2000yi,Guasch:2001wv}.   The
difference between $\delta h_f$ and  $\Delta h_f$ is that $\delta h_f$
is  a  radiative correction  to  the  supersymmetric $h_f^{\rm  MSSM}$
coupling  of   up-quarks,  down-quarks  and  leptons.    The  coupling
$\Delta h_f$, instead,  is a loop-induced coupling of  the fermions to
the Higgs  doublet to which they  do not  couple in  the supersymmetric
limit.  Therefore, below the scale  $M_S$ the theory becomes a general
2HDM, with up-quarks  coupled to $\Phi_2$ and  down-quarks and leptons
coupled     to      $\Phi_1$,     with     couplings      given     by
$h_{f}^{\rm MSSM} ( 1 +  \delta_f)$, respectively, but with additional
loop-induced  couplings $\Delta  h_f$ to  the other  Higgs doublet.
The couplings  $h_f^{\rm 2HDM}$ are  the combinations of  these Yukawa
couplings related to the running fermion masses in the same way
as in a Type-II 2HDM.   Notice that in  the  present  approach,  we  treat  the  loop-induced
couplings $\Delta h_f$ as small departures from a Type-II 2HDM. Hence,
we are  working in  a Type-II  approximation to  a general  2HDM.

The  RGEs for  the 2HDM  used for  $M_S >  Q >  M_H$ are  described in
Appendices~\ref{RGE:2HDM1loop} and \ref{RGE:2HDM2loop}.     At    the    heavy    Higgs    threshold
$M_H  \equiv   M_{H^\pm}$,  the  following  matching   conditions  are
employed:
\begin{eqnarray}
&&
\lambda\ =\ \frac{\left(M_{H_1}^{\rm EP}\right)^2}{v^2}\:
-\: \frac{1}{16}\kappa(g^2+g'^2)^2s_{4\beta}^2
\, ,
\nonumber \\[2mm] &&
h_t^{\rm 2HDM}\ =\ \frac{y_t}{s_\beta}\ , \qquad
h_b^{\rm 2HDM}\ =\ \frac{y_b}{c_\beta}\ , \qquad
h_\tau^{\rm 2HDM}\ =\ \frac{y_\tau}{c_\beta} \ ,
\label{eq:2HDM}
\end{eqnarray}
where $M_{H_1}^{\rm EP}$  denotes the effective potential  mass of the
lightest neutral Higgs boson calculated  in the limit of zero external
momentum $s=0$.  In  the above, we have ignored the  small effects due
to scheme  conversion from  dimensional regularization  to dimensional
reduction~\cite{Draper:2013oza}.   In practice,  while evaluating  the
evolution   of   the  gauge,   Yukawa   and   quartic  couplings,   at
$M_H < Q <  M_S$, we have assumed an effective  Type-II 2HDM, in which
the  Yukawa couplings  are given  by $h_f^{\rm  2HDM}$ with  the 
matching condition, Eq.~(\ref{eq:2HDM}) given  at the scale  $M_H$.  As already  mentioned above,
this  amounts   to  an  approximate  treatment   of  the  loop-induced
$\Delta h_f$ effects on the computation  of the Higgs boson masses and
mixing angles.

At scales below the heavy Higgs scale $M_H$, the only physical degrees of freedom are the 
SM ones.   The  RGEs for the  SM used for  $Q <
M_H$ are described in Appendix~A. We define the SM Higgs potential as
$$
V(\Phi)\ =\ -\,\frac{m^2}{2}|\Phi|^2\: +\: \frac{\lambda}{2}|\Phi|^4\;,
$$
with $\Phi=(0,(v+h)/\sqrt{2})^T$ and $\lambda=m^2/v^2$.  Note that the
quartic coupling  of $\Phi$ is  defined with a factor  (-2) difference
compared to the quartic couplings of $\Phi_{1,2}$ in Eq.~(\ref{LV}).
In order to compare with  the experimental results, it is important to
define  the   SM  boundary  conditions   for  the  gauge,  Yukawa and quartic
couplings. In our work we use~\cite{Buttazzo:2013uya}
\begin{eqnarray}
\label{eq:mtcouplings}
y_t &=& 0.93697 + 0.00550 \left( \frac{m_t^{\rm pole}}{\rm GeV} - 173.35 \right) 
- 0.00042 \frac{\alpha_s(M_Z) - 0.1184}{0.0007} \, ,
\nonumber \\[2mm]
g_s &=& 1.1666 + 0.00314 \frac{\alpha_s(M_Z) - 0.1184}{0.0007} - 
0.00046 \left( \frac{m_t^{\rm pole}}{\rm GeV} - 173.35 \right) \, ,
\nonumber \\[2mm] 
g'&=&0.3587\,, \ \ \
g=0.6483\,, \ \ \
y_b=0.0156\,, \ \ \
y_\tau=0.0100 \, .
\end{eqnarray}
The pole mass-squared of the lightest Higgs boson is then given by~\cite{Draper:2013oza}:
\begin{eqnarray}
   \label{mtpole}
&& \hspace{-0.9cm}(M_{H_1}^{\rm pole})^2\  =\
\lambda (m_t^{\rm pole})\,  v^2 (m_t^{\rm pole})  
\nonumber \\ &&
\qquad+\ \kappa \Bigg\{
3y_t^2 (4\overline{m}_t^2 - m_h^2) B_0(m_h^2,\overline{m}_t^2, \overline{m}_t^2) - 
\frac{9}{2} \lambda m_h^2 
\left[2 - \frac{\pi}{\sqrt{3}} - \log \frac{m_h^2}{Q_{\rm RG}^2} \right] 
\nonumber \\ &&
\qquad -\ \frac{v^2}{4} \left[ 3g^4 - 4\lambda g^2 + 4\lambda^2 \right] 
B_0(m_h^2,M_W^2, M_W^2) 
\nonumber \\ &&
\qquad -\ \frac{v^2}{8} \left[ 3(g^2 + g'^2)^2 - 4\lambda (g^2 + g'^2) 
+ 4\lambda^2 \right] B_0(m_h^2,M_Z^2, M_Z^2)
\nonumber \\ &&
\qquad +\ \frac{1}{2} g^4 \left[ g^2 - \lambda \left( \log\frac{M_W^2}{Q_{\rm RG}^2} - 1 \right)
\right] + \frac{1}{4} (g^2 + g'^2) \left[ (g^2 + g'^2) 
- \lambda \left( \log\frac{M_Z^2}{Q_{\rm RG}^2} - 1 \right) \right]
\Bigg\} \, , \nonumber \\
\end{eqnarray}
where $\overline{m}_t =y_tv/\sqrt{2}$ and
$m_h^2=\lambda (m_t^{\rm pole})\, v^2 (m_t^{\rm pole})$. We take the 
renormalization group scale 
$Q_{\rm RG}=m_t^{\rm pole}$, and the function $B_0$ used in
(\ref{mtpole}) is defined in~\cite{Carena:2001fw}.

\section{Numerical Results for the MSSM Higgs Sector}

We first  illustrate the  effects of  the RG running  in the  range of
scales       $Q       >       M_H$      using       a      specific
 scenario with universal  SUSY parameters 
fixed to be $1$ TeV:
\begin{equation}
   \label{eq:SCN} 
\mu\ =\ M_{1,2,3}\ =\ M_{\tilde{Q}_3,\tilde{U}_3,\tilde{D}_3,
\tilde{L}_3,\tilde{E}_3}\ =\ A_{t,b,\tau}\ =\ 1~{\rm TeV}\;, \qquad
\rho_{\tilde{Q},\tilde{U},\tilde{D},\tilde{L},\tilde{E}}\ =\ 1\;,
\end{equation}
where   $\rho_{\tilde{Q}}\  =\   M_{\tilde{Q}_{1,2}}/M_{\tilde{Q}_3}$,
$\rho_{\tilde{U}}\  =\ M_{\tilde{U}_{1,2}}/M_{\tilde{U}_3}$,  etc, and
we have assumed no hierarchy between the three generations of sfermion
masses.

\begin{figure}[t!]
\begin{center}
\includegraphics[width=12cm]{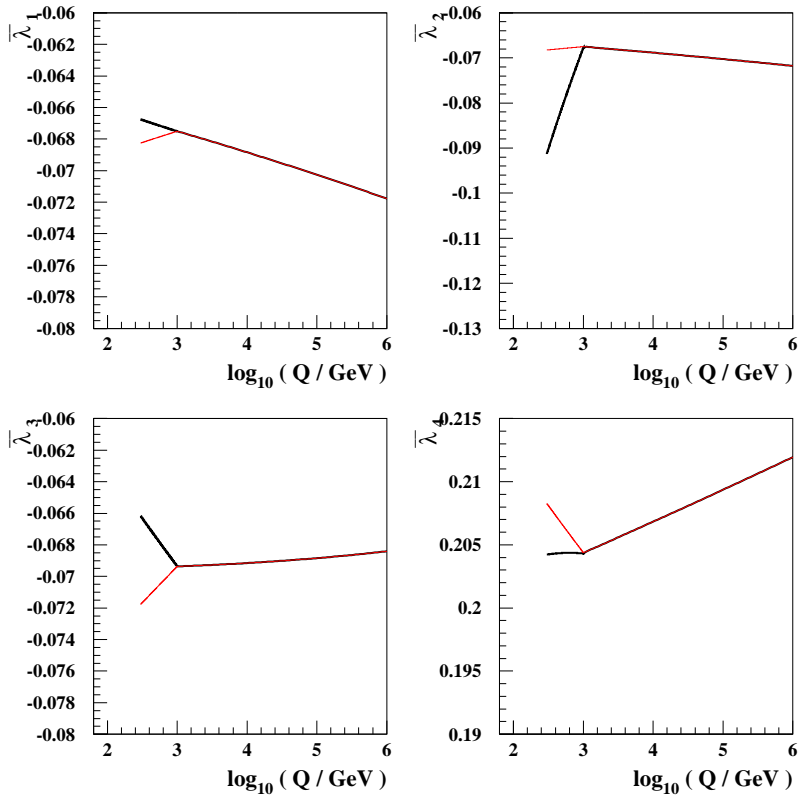} 
\end{center}
\vspace{-0.5cm}
\caption{\it The black lines show the one-loop running of the 2HDM
  quartic couplings for $Q<10^6$ GeV, assuming $M_{H^\pm}=300$ GeV and
  $\tan\beta=5$.  The other input parameters are given in
  Eq.~(\ref{eq:SCN}).  The thin red lines show the running of
  $-(g^2+g'^2)/8$, $-(g^2-g'^2)/4$, and $g^2/2$ in the panels for
  $\bar\lambda_{1,2}$, $\bar\lambda_{3}$, and $\bar\lambda_{4}$,
  respectively.  }
\label{fig:q1234.enl}
\end{figure}

Figure~\ref{fig:q1234.enl} illustrates with black lines the one-loop
running of the 2HDM quartic couplings up to $Q = 10^6$~GeV in the
above scenario, Eq.~(\ref{eq:SCN}), and compares them with the running
of the corresponding combinations of electroweak gauge couplings (red
lines).  Since there is a single SUSY-breaking scale of 1~TeV and
hence a single threshold, the couplings are matched at this scale, and
the red and black lines diverge as $Q$ decreases from $M_S = 1$~TeV to
$M_H = M_{H^\pm} = 300$~GeV.  Above $M_S$ the RG evolution of the
quartic couplings (black lines) are the same as those of the
corresponding combinations of electroweak gauge couplings (red lines),
i.e., the red and black lines lie on top of each other.  This provides
a non-trivial consistency check for the correctness of our results.

\begin{figure}[t!]
\begin{center}
\includegraphics[width=12cm]{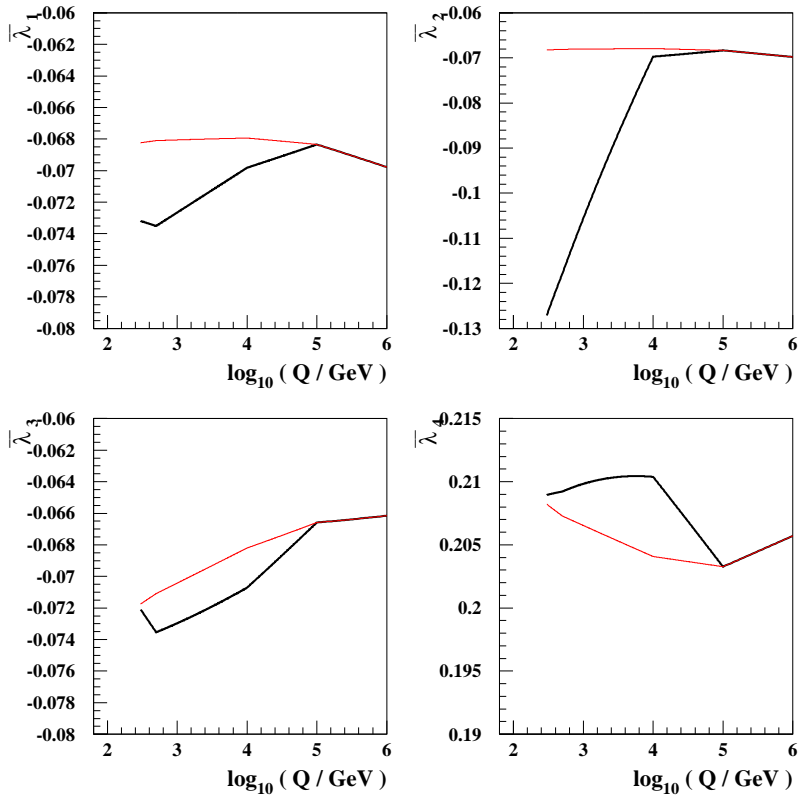} 
\end{center}
\vspace{-0.5cm}
\caption{\it The  same as in Fig.~\ref{fig:q1234.enl}, but for an
  hierarchical scenario with $\rho = 10$ and
  input parameters given in Eq.~(\ref{eq:SCN2}).}
\label{fig:q1234.2.enl}
\end{figure}

In the same context, it is important to comment that, for hierarchical
scenarios         with        $\rho         >        1$,         where
$\rho                \equiv                 {\rm                max}\,
(\rho_{\tilde{Q},\tilde{U},\tilde{D},\tilde{L},\tilde{E}})$,
the proper matching  conditions should be imposed at  the highest soft
SUSY-breaking  scale $M'_S  = \rho  M_S$,  rather than  $M_S$.  As  an
illustrative  example,  we  consider  another scenario  with  various
different values for the soft SUSY breaking parameters:
\begin{eqnarray}
&&
\mu\ =\ 500 ~{\rm GeV}\,,\qquad M_1\ =\ 100 ~{\rm GeV}\,,\qquad M_2\ =\ 200 ~{\rm GeV}\,,\qquad
M_3\ =\ 2\,i ~{\rm TeV}\, ,\qquad \nonumber\\
 &&
m_{\tilde{Q}_3,\tilde{U}_3,\tilde{D}_3,
\tilde{L}_3,\tilde{E}_3}\ =\ 10 ~{\rm TeV}\,,\qquad  A_{t,b,\tau}\ =\ 1\,i ~{\rm TeV} \,,\qquad
\rho_{\tilde{Q},\tilde{U},\tilde{D},\tilde{L},\tilde{E}}\ =\ 10\,.
   \label{eq:SCN2}
\end{eqnarray}
In this scenario,  the matching conditions are imposed  at the highest
soft SUSY-breaking scale $M_S'=10^5$ GeV instead of $M_S=10^4$ GeV, as
illustrated  in  Figure~\ref{fig:q1234.2.enl}.   We observe  that  the
running  between $M_S$  and $M_S'$  changes the  size of  the quartic
couplings by  an amount of  $\sim 2$\% for $\rho=10$,  which
results in  a less than $1$~GeV increase in  the mass prediction for  the $H_1$
boson.   Even though  such changes  may  not appear  too significant
for  scenarios with mass
spectrum hierarchies  of $\rho  \stackrel{<}{{}_\sim} 10$,  they are nevertheless
accurately described  within our  multi-threshold RG approach  that we
follow here for  the computation of the Higgs-boson  masses and mixing
angles.

\begin{figure}[t!]
\vspace{-0.5cm}
\begin{center}
\includegraphics[width=9.5cm]{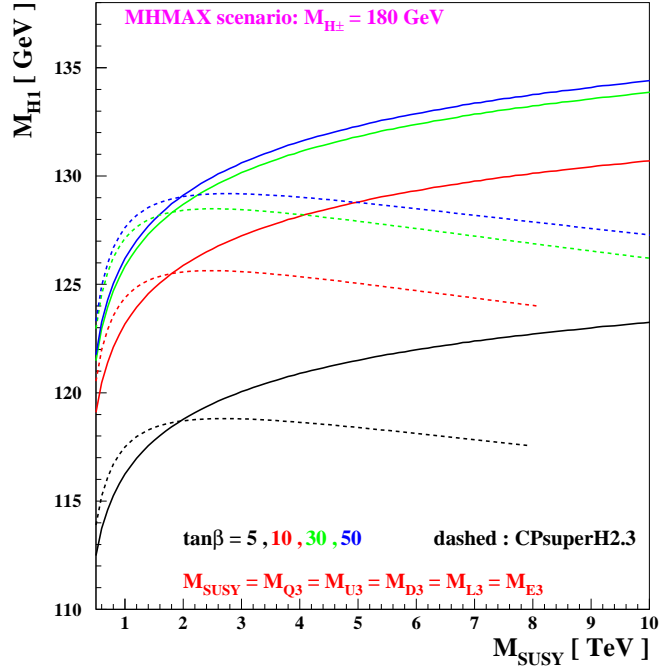}
\end{center}
\vspace{-0.5cm}
\caption{\it The lightest Higgs mass calculated in the MHMAX scenario
  using {\tt CPsuperH3.0} (solid lines) and {\tt CPsuperH2.3} (dashed
  lines) as functions of
  $M_S=M_{Q3} = M_{U3} = M_{D3} = M_{L3} = M_{E3}$ for $M_{H^\pm}=180$
  GeV, $\mu=200$ GeV and various values of $\tan \beta$.  }
\label{fig:mh1.mhmax}
\end{figure}

\begin{figure}[t!]
\begin{center}
\includegraphics[width=8.1cm]{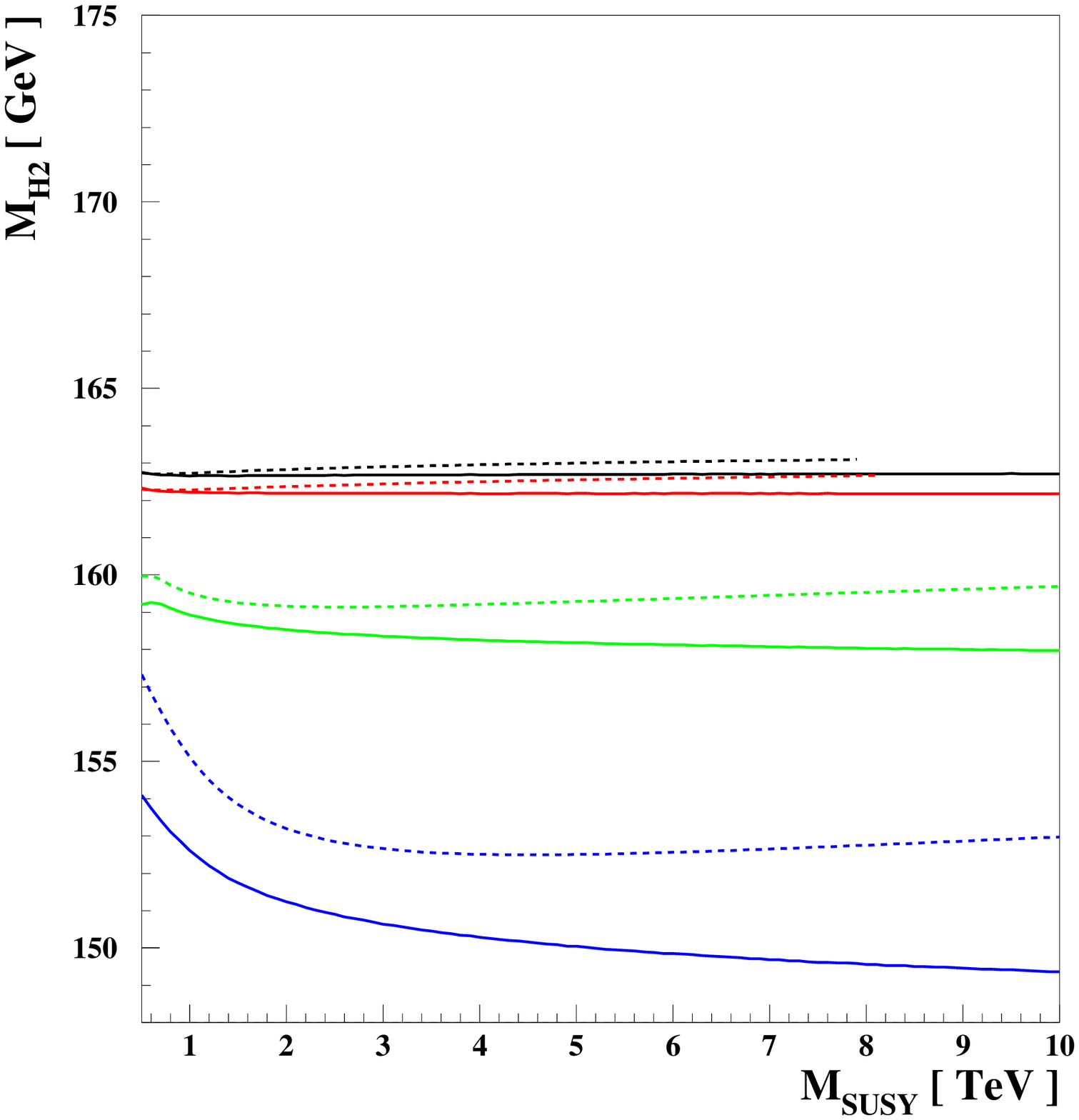}
\includegraphics[width=8.1cm]{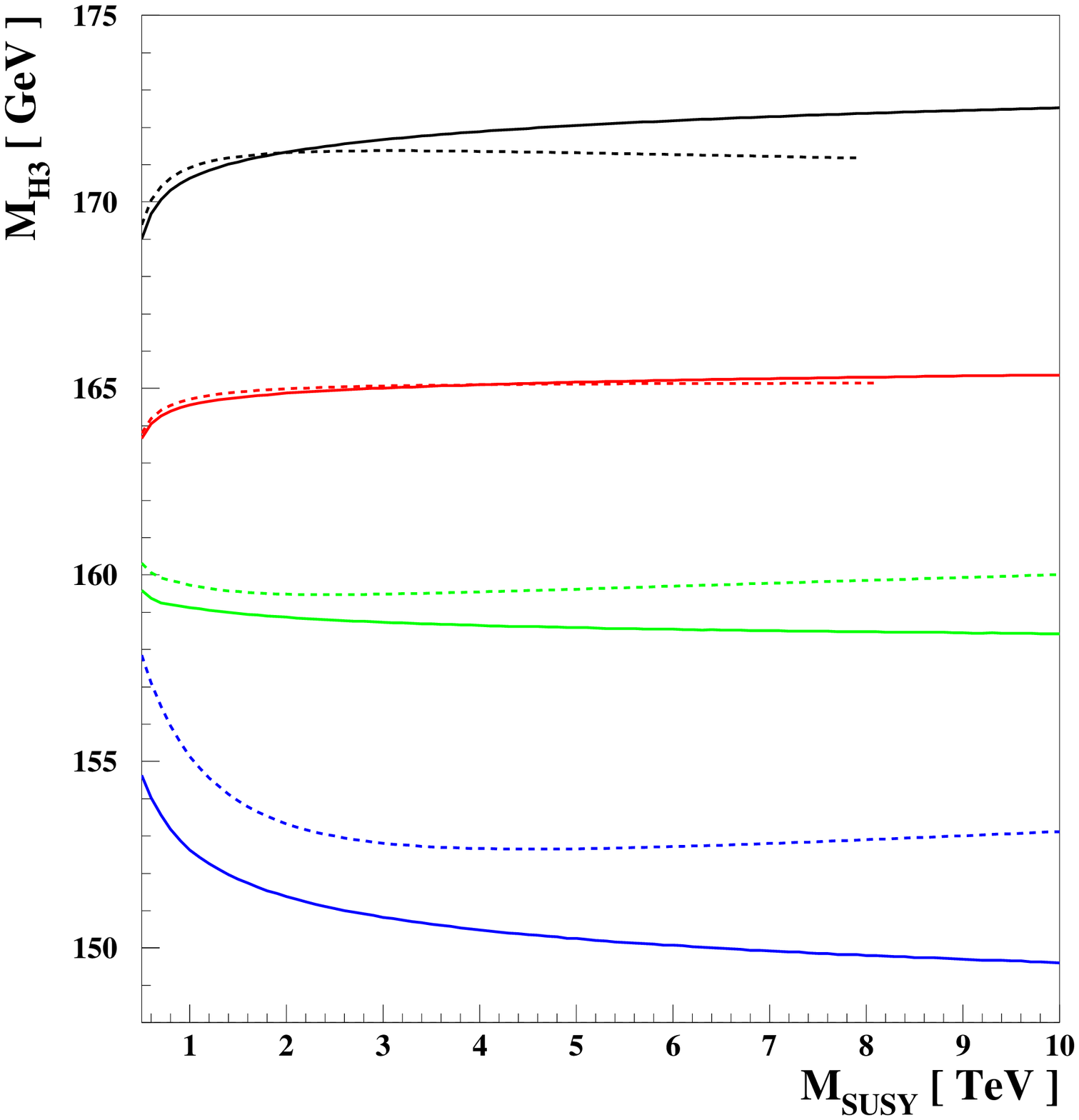}
\end{center}
\vspace{-0.5cm}
\caption{\it The two heavier neutral Higgs masses $M_{H_{2,3}}$
  calculated in the MHMAX scenario using {\tt CPsuperH3.0} (solid
  lines) and {\tt CPsuperH2.3} (dashed lines), for the same parameter
  choices as in Figure~\protect\ref{fig:mh1.mhmax}.  }
\label{fig:mh23.mhmax}
\end{figure}

Next, we compare some results of {\tt CPsuperH3.0} with the
corresponding results of {\tt CPsuperH2.3} in the MHMAX
scenario\cite{Carena:2002qg,Carena:2013ytb}, where
$X_t = \sqrt{6} M_S$ and $\mu = 200$~GeV.  In order to isolate the
effects of the running in the range {$M_H<Q<M_S$, where the effective
  2HDM description is valid}, we consider examples with
$M_{H^\pm} \simeq m_t^{\rm pole}$.  Figure~\ref{fig:mh1.mhmax}
compares calculations of the lightest Higgs mass $M_{H_1}$, and
Figure~\ref{fig:mh23.mhmax} shows results for the two heavier neutral
Higgs bosons $H_{1,2}$, {for $M_{H^\pm}=180$~GeV}.  In order to
isolate the effects of the resummation of logarithms associated with
the RG effects, we modified {\tt CPsuperH2.3}, setting
$m_t(m_t^{\rm pole}) = 162.88$~GeV as obtained using
Eq.~(\ref{eq:mtcouplings}), instead of the value obtained from the
one-loop relation between the pole and running masses:
$ m_t(m_t^{\rm pole})=m_t^{\rm pole}/ [1+ 4\alpha_s(m_t^{\rm
  pole})/(3\pi)] \simeq$165.5~GeV,
that was the standard value in {\tt CPsuperH2.3}.  The lower value of
the running top quark mass we use is based on higher-order loop
corrections to the relation between the pole and running
masses~\cite{Buttazzo:2013uya},\cite{Melnikov:2000qh} and, as stressed
before, it is used as the standard value for {\tt CPsuperH3.0}.

From Fig.~\ref{fig:mh1.mhmax} 
we see that in the MHMAX scenario,  the mass of the lightest Higgs boson
calculated using {\tt  CPsuperH3.0} is $\sim 1$ GeV  smaller than that
obtained using {\tt CPsuperH2.3} for $M_S = M_{\tilde{Q}_3} = M_{\tilde{U}_3} = M_{\tilde{D}_3} =
M_{\tilde{L}_3} =  M_{\tilde{E}_3} = 1$ TeV  \footnote{Here we ignore  the small difference of
  $M_S$ from that defined  in Eq.~(\ref{defineMS}),  but this difference
  is  taken  into account in all the  numerical results presented
  in this work.}.
This $\sim 1$ GeV difference may be attributed to the use of $h_t^{\rm
  2HDM}$ in {\tt CPsuperH3.0} in the running to low energies.  Namely,
whereas  in {\tt  CPsuperH2.3} the  top Yukawa  coupling  appearing in
$\lambda_2^{(1),(2)}$    given   by    Eqs.~(3.4)   and    (3.10)   of
Ref.~\cite{Carena:2000yi} includes the  threshold corrections, we have
not  included these  corrections in  {\tt CPsuperH3.0}.   These Yukawa
thresholds  are still  included  in the  relevant  computation of  the
threshold  corrections to  the quartic  couplings, which  lead  to the
asymmetry     between    positive     and    negative     values    of
$X_t=A_t-\mu^*/\tan\beta$  in the CP-conserving  limit of  the theory.
This  small   difference  between  the  {\tt   CPsuperH3.0}  and  {\tt
  CPsuperH2.3} is rapidly compensated  by RG effects and, as expected,
the mass  difference changes sign when  $M_S \sim 2$~TeV,  and the new
calculation of  $M_{H_1}$ is larger than  the old one  by $\sim 5$~GeV
when $M_S \sim  8$~TeV, the difference being only  weakly dependent on
$\tan  \beta$.  We  do  not  show results  for  {\tt  CPsuperH2.3}  at
$\tan\beta = 5$ and $10$ and $M_S > 8$~TeV, since this program becomes
unstable for that region of parameters.

Figure~\ref{fig:mh23.mhmax}  shows  corresponding  comparisons of  the
masses of the  two heavier neutral Higgs bosons  $M_{H_{2,3}}$ for the
same parameter values.  We see that the differences are very small for
$\tan \beta  = 5$ and $10$,  but increase for larger  $\tan \beta$ and
larger $M_S$, reaching  $\sim 4$~GeV for $\tan \beta =  50$ and $M_S =
10$~TeV.

\begin{figure}[t!]
\begin{center}
\includegraphics[width=9.5cm]{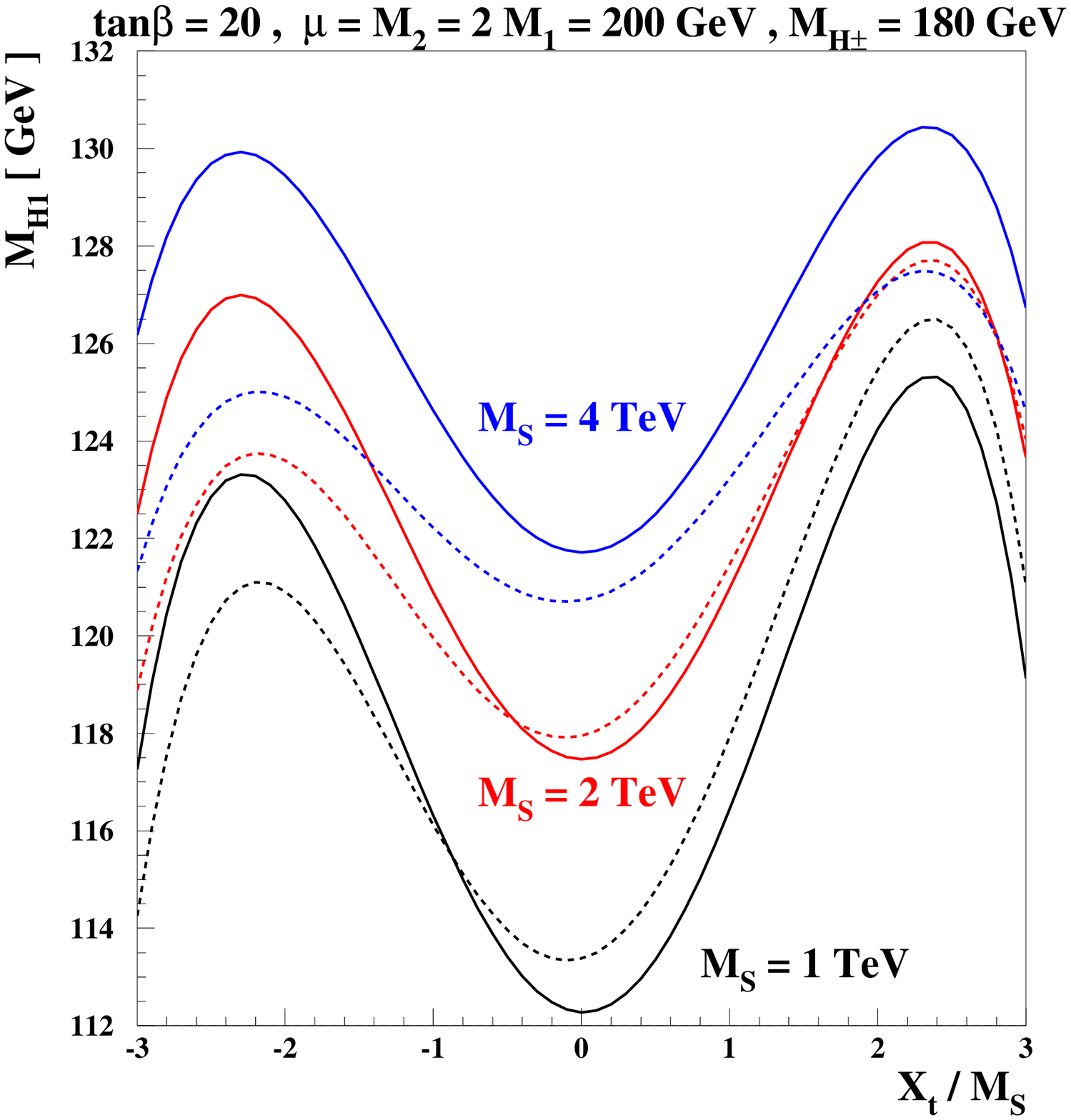}
\end{center}
\vspace{-0.5cm}
\caption{\it The lightest Higgs mass calculated as a function of
  $X_t/M_S$ using {\tt CPsuperH3.0} (solid lines) and {\tt
    CPsuperH2.3} (dashed lines) for $M_{H^\pm}=180$ GeV,
  $\mu=M_2 = 2 M_1 = 200$~GeV and $\tan \beta = 20$ for $M_S = 1$~TeV
  (black lines), 2~TeV (red lines) and 4~TeV (blue lines).  }
\label{fig:mh1.xt0}
\end{figure}

\begin{figure}[t!]
\vspace{-0.5cm}
\begin{center}
\includegraphics[width=9.5cm]{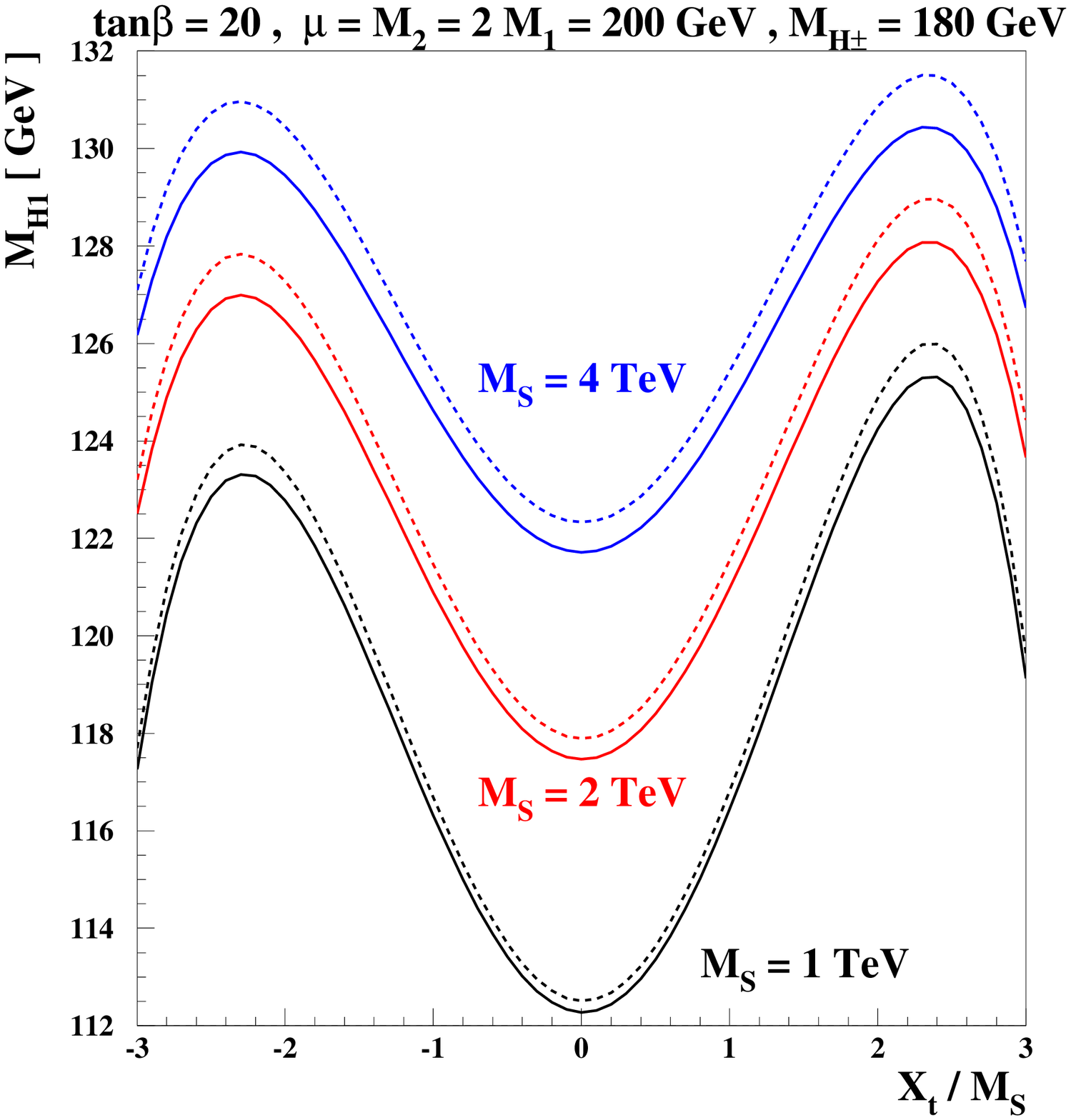}
\end{center}
\vspace{-0.5cm}
\caption{\it
The lightest Higgs mass calculated as a function of $X_t/M_S$
including (solid lines) and without including (dashed lines) 
the two-loop 2HDM contributions to the RGEs
for $M_{H^\pm}=180$ GeV, $\mu=M_2 = 2 M_1 = 200$~GeV and $\tan \beta = 20$
for $M_S = 1$~TeV (black lines), 2~TeV (red lines)
and 4~TeV (blue lines).
}
\label{fig:mh1.1.2.loops}
\end{figure}

Figure~\ref{fig:mh1.xt0} shows another comparison of calculations of
$M_{H_1}$ made using {\tt CPsuperH3.0} (solid lines) and {\tt
  CPsuperH2.3} (dashed lines), this time as a function of $X_t/M_S$
for $M_S = 1$~TeV (black lines), 2~TeV (red lines) and 4~TeV (blue
lines).  These calculations were made assuming $M_{H^\pm}=180$ GeV,
$\mu=M_2 = 2 M_1 = 200$~GeV and $\tan \beta = 20$.  We see that the
differences in $M_{H_1}$ are again small, i.e., $\lsim 1$~GeV, for
most values of $X_t/M_S$, though rising to $\sim 2$~GeV for
$X_t/M_S \sim -2$. The results of {\tt CPsuperH3.0} are in agreement with
those obtained in Ref.~\cite{Lee:2015uza}. Figure~\ref{fig:mh1.1.2.loops} compares
calculations of $M_{H_1}$ made within {\tt CPsuperH3.0} using the
two-loop 2HDM RGEs (solid lines) and the one-loop 2HDM RGEs (dashed
lines), again as a function of $X_t/M_S$ for $M_S = 1$~TeV (black
lines), 2~TeV (red lines) and 4~TeV (blue lines). We see that the full
two-loop results are generally smaller than those in the one-loop
approximation by $< 1$~GeV, providing an encouraging estimate of their
reliability.

\begin{figure}[t!]
\vspace{-0.5cm}
\begin{center}
\includegraphics[width=8.1cm]{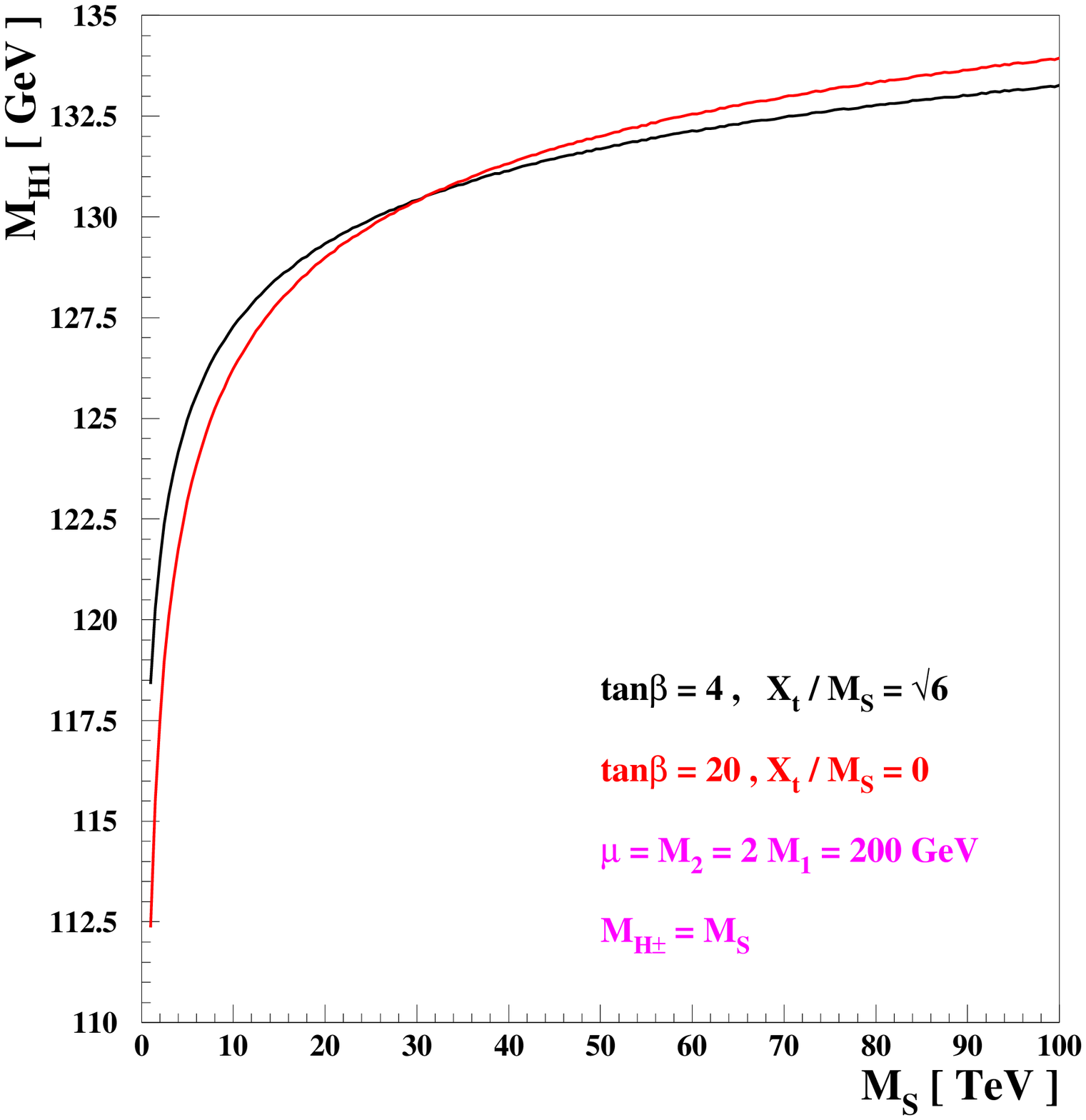}
\includegraphics[width=8.1cm]{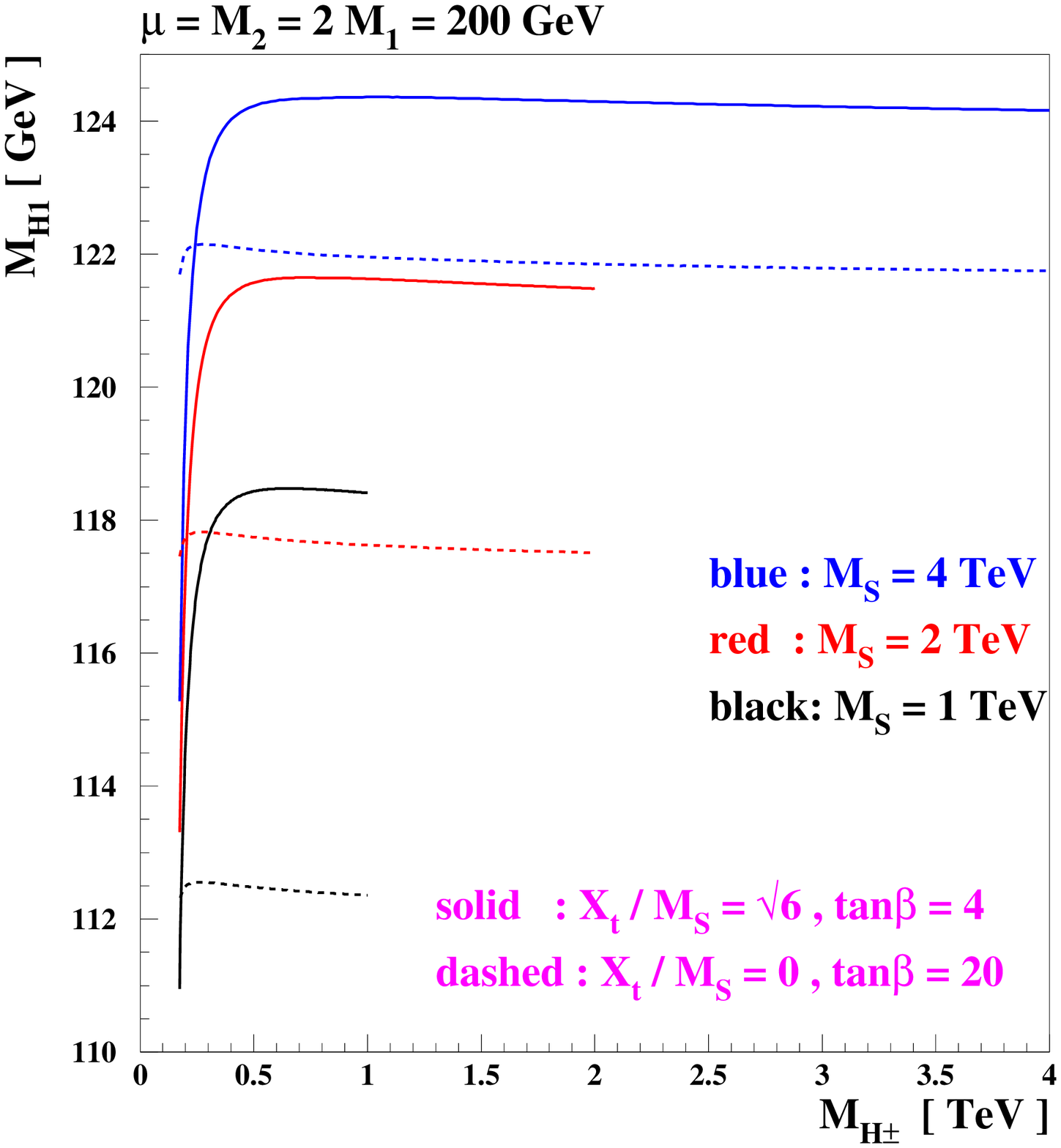}
\end{center}
\vspace{-0.5cm}
\caption{\it Left panel: Calculations of $M_{H_1}$ made using {\tt
    CPsuperH3.0} for $M_{H^\pm} = M_S \le 100$~TeV, assuming
  $\tan \beta = 4$ and $X_t/M_S = \sqrt{6}$ (black line) and
  $\tan \beta = 20$ and $X_t/M_S = 0$ (red line) both for the case
  $\mu=M_2 = 2 M_1 = 200$~GeV. Right panel: calculations of $M_{H_1}$
  as a function of $m_{H^\pm}$ made using {\tt CPsuperH3.0} over the
  range $M_{H^\pm} \in [m_t, M_S]$ for $M_S = 1, 2$ and 4~TeV (black,
  red and blue lines, respectively), also for the two cases
  $X_t/M_S = \sqrt{6}$ and $\tan \beta =4$ (solid lines)  and $X_t/M_S = 0$ and
  $\tan \beta =20$ (dashed lines) and $\mu=M_2 = 2 M_1 = 200$~GeV. These results can
  be compared with those of~\cite{Draper:2013oza}.}
\label{fig:mh1.fig2}
\end{figure}

Figure~\ref{fig:mh1.fig2}  shows some  results from  {\tt CPsuperH3.0}
for  some larger  values of  $M_S$ and  $M_H =  M_{H^\pm}$ where  {\tt
  CPsuperH2.3}  would have  been inapplicable.   The left  panel shows
$M_{H_1}$  for   $M_{H^\pm}  =  M_S   \le  100$~TeV  for   two  cases,
$\tan  \beta  =   4$  and  $X_t/M_S  =  \sqrt{6}$   (black  line)  and
$\tan \beta  = 20$  and $X_t/M_S  = 0$  (red line)  both for  the case
$\mu=M_2     =     2     M_1      =     200$~GeV     considered     in
Figure~\ref{fig:mh1.xt0}. There  is no  2HDM running effects  in these
plots, only  the effects of  SM running up  to the common  new physics
threshold.  Even allowing  for an uncertainty of about  3~GeV in these
calculations, values  of $M_S =  M_{H^\pm}> 15$~TeV lead to  values of
the  lightest Higgs  mass  $M_{H_1}$ that  are  incompatible with  the
measured values  at the  LHC, so  the extension of  the $M_S$  axis to
100~TeV is largely for illustrative purposes.

The right panel of Figure~\ref{fig:mh1.fig2} shows some other results
for larger values of $M_{H^\pm}$ over the full range $[m_t, M_S]$ for
$M_S = 1, 2$ and 4~TeV (black, red and blue lines, respectively) in
the same two cases $X_t/M_S = \sqrt{6}$,  $\tan \beta =4$ and
$X_t/M_S = 0$, $\tan \beta =20$, considered previously, again with
$\mu=M_2 = 2 M_1 = 200$~GeV. {For these values of $\tan\beta$, the
  measured values of the Higgs mass is consistent with $M_S = 4$~TeV
  in the $X_t/M_S = \sqrt{6}$, $\tan \beta =4$
  case}~\footnote{{The results in the left and right panes of Figure~\ref{fig:mh1.fig2} should be compared to
    the ones in the  lower-left frames of Figs.~1 and 2, and of Fig. 6 in
    Ref.~\cite{Draper:2013oza}}}.

\vfill\eject

\section {CP-Violating Heavy Higgs Scenarios}

We  now  consider various  \textcolor{red}{CP}\textcolor{green}{X}{\it
  4}\textcolor{blue}{LHC} benchmark scenarios for showcasing the effect
of CP  violation in  the MSSM  heavy Higgs  sector and  their possible
signatures.     We    assume     a    common     CP-violating    phase
$\Phi_A={\rm arg}(A_t)={\rm arg}(A_b)={\rm arg}(A_\tau)$, set
\begin{eqnarray}
|A_{t,b,\tau}|\ =\ \mu\ =\ 2\,M_S\; ,
\end{eqnarray}
with  $M_2=2\,M_1=200$  GeV and  $M_3=2$  TeV,  and vary  $\tan\beta$,
$M_{H^\pm}$,  and $M_S$.   We do  not include  gaugino phases  in this
analysis, as  they enter the  Higgs sector only through  the threshold
corrections    to   the   MSSM    top-,   bottom-,    and   tau-Yukawa
couplings. Instead, we include the CP-conserving leading-log
enhanced   contributions  due   to  gauginos   to   the  self-energies
$\Pi^{\pm,S,P}$.   In  our \textcolor{red}{CP}\textcolor{green}{X}{\it
  4}\textcolor{blue}{LHC} scenarios, since  we  fix  $M_2=2
M_1=200$~  GeV and  $M_3=2$~TeV, and  do  not increase  them as  $M_S$
increases, the gaugino phase effects are relatively insignificant.

\begin{figure}[t!]
\begin{center}
\includegraphics[width=9.5cm]{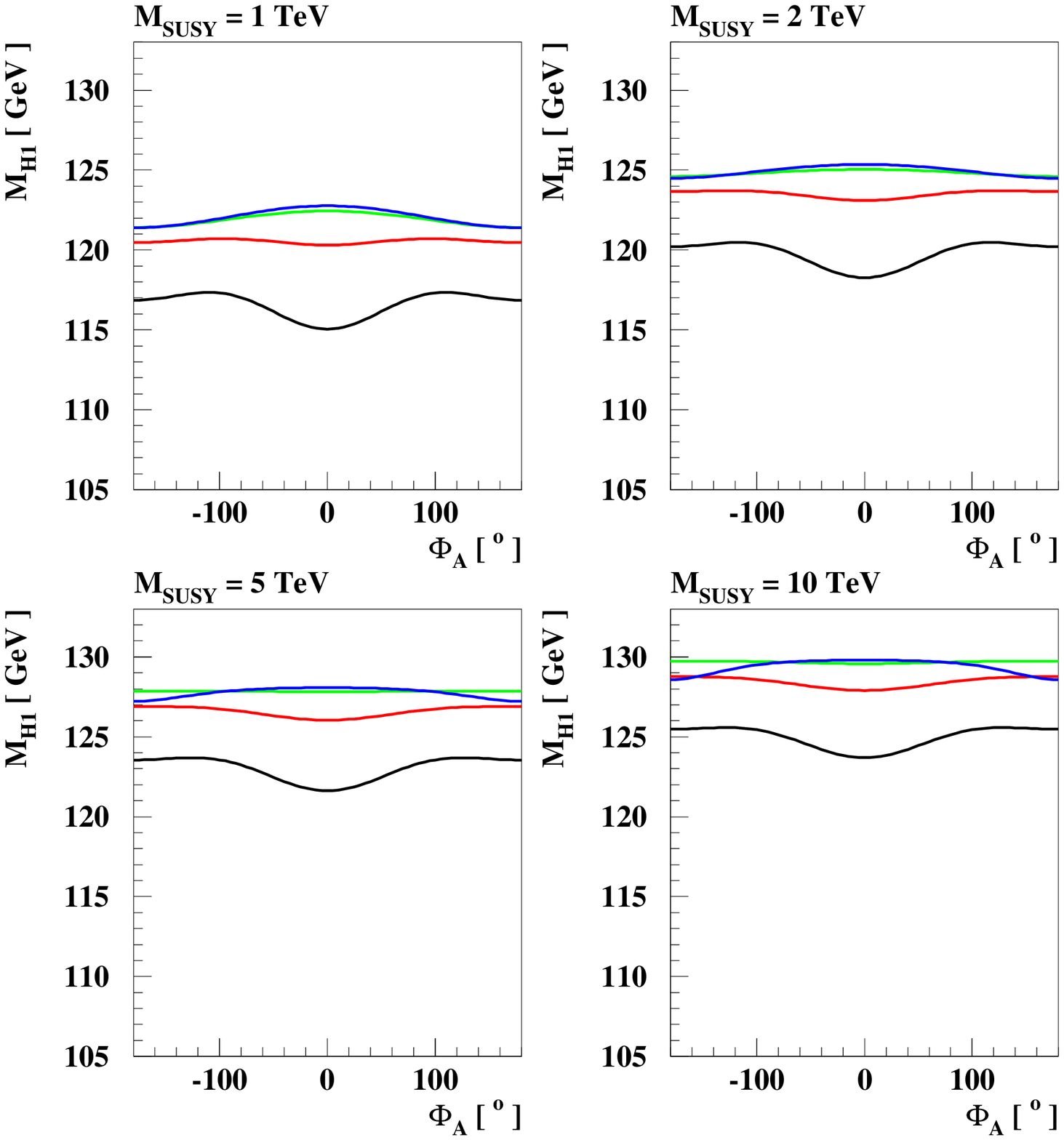}
\end{center}
\vspace{-0.75cm}
\caption{\it Values of $M_{H_1}$ in CP-violating scenarios for
  $M_{H^\pm}=500$ GeV and $M_S=1$ TeV (upper left), $M_S=2$ TeV (upper
  right), $M_S=5$ TeV (lower left), $M_S=10$ TeV (lower right).  The
  black, red, green and blue lines are for $\tan\beta=5, 10, 30$ and
  $50$, respectively.  }
\label{fig:cpx.mh1}
\end{figure}

We first present in Figure~\ref{fig:cpx.mh1} results for $M_{H_1}$ as
a function of $\Phi_A$ for the representative choices $M_{H^\pm}= 500$
GeV, $M_S=1, 2, 5, 10$ TeV and $\tan\beta=5, 10, 30, 50$. The changes
in $M_{H_1}$ as $\Phi_A$ varies are small in general, namely
$\lsim 3$~GeV for $\tan \beta = 5$ and less for larger $\tan \beta$.
Nevertheless, these variations are potentially significant, as there
are parameter choices that would be excluded (in the sense of yielding
values of $M_{H_1}$ more than 3~GeV different from the measured value)
for $\Phi_A = 0$ that would be allowed for $\Phi_A \ne 0$, e.g.,
$M_S = 5$~TeV and $\tan \beta = 5$. Conversely, there are cases where
$\Phi_A = 0$ would be allowed, but $\Phi_A \ne 0$ would be disallowed,
e.g., $M_S = 10$~TeV and $\tan \beta = 10$.  {The change in the
  lightest Higgs mass $M_{H_1}$ at lower values of $\tan\beta$ can be
  understood from the change in the modulus of
  $X_t = A_t - \mu^*/\tan\beta$, which governs the one-loop threshold
  corrections to the low-energy Higgs quartic coupling. It reaches a
  maximum when $|X_t|/M_S \simeq 2.4$, and becomes less significant as
  $\tan\beta$ increases.  In addition to this change, there are
  two-loop effects governed by the relative phase of $A_t$ and $M_3$,
  which in the CP-conserving case tend to decrease (increase)
  $M_{H_1}$ for negative (positive) values of $A_t M_3^*$, which
  explains why, for large values of $\tan\beta$, for which the variation of $|X_t|$ is small, the maximum value of
  $M_{H_1}$ occurs for $\Phi(A_t) = 0$. }

\begin{figure}[t!]
\vspace{-1.0cm}
\begin{center}
\includegraphics[width=9.5cm]{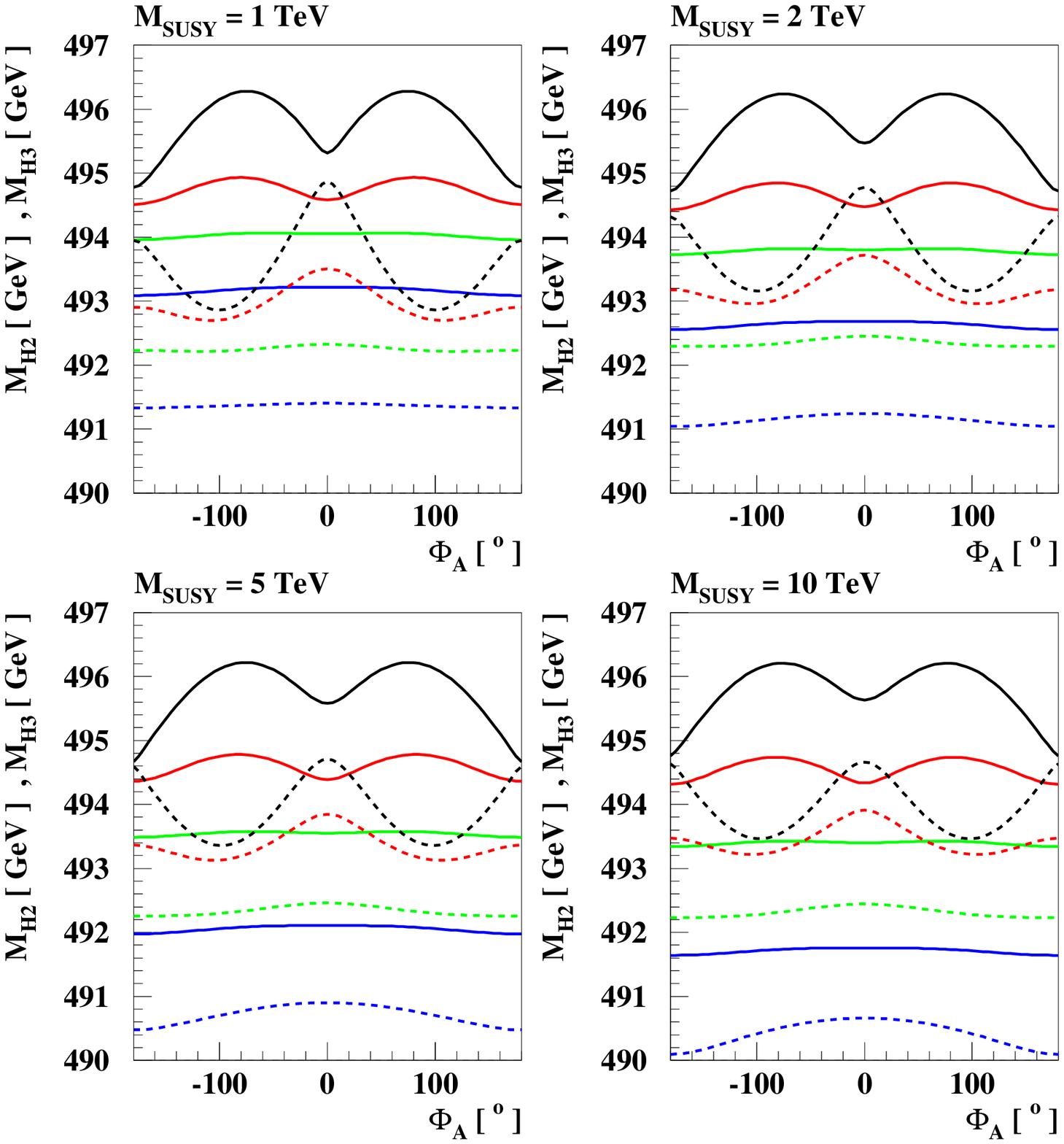}
\end{center}
\vspace{-0.5cm}
\caption{\it   
Values of $M_{H_3}$ (solid lines) and $M_{H_2}$ (dashed lines)
in CP-violating scenarios for $M_{H^\pm}=500$ GeV and
$M_S=1$ TeV (upper left),
$M_S=2$ TeV (upper right),
$M_S=5$ TeV (lower left),
$M_S=10$ TeV (lower right).
The black, red, green and blue lines are for
$\tan\beta=5, 10, 30$ and $50$, respectively.
}
\label{fig:cpx.mh23}
\end{figure}

Figure~\ref{fig:cpx.mh23} displays the corresponding values of
$M_{H_3}$ (solid lines) and $M_{H_2}$ (dashed lines) for
$M_{H^\pm}=500$ GeV and the same values of $M_S$ and $\tan\beta$ as in
Figure~\ref{fig:cpx.mh1}. We see that in general the mass difference
$M_{H_3} - M_{H_2}$ is minimized in the CP-conserving cases
$\Phi_A = 0, 180^\circ$, where it is $\lsim 1$~GeV, and maximized when
$\Phi_A = \pm 90^\circ$, where it may be $\sim 3$~GeV.  Thus, a
measurement of the ${H_3} - {H_2}$ mass difference could be an
indirect diagnostic tool indicative of CP violation, even if the
latter is not directly observable.

\begin{figure}[t!]
\vspace{-1.0cm}
\begin{center}
\includegraphics[width=9.5cm]{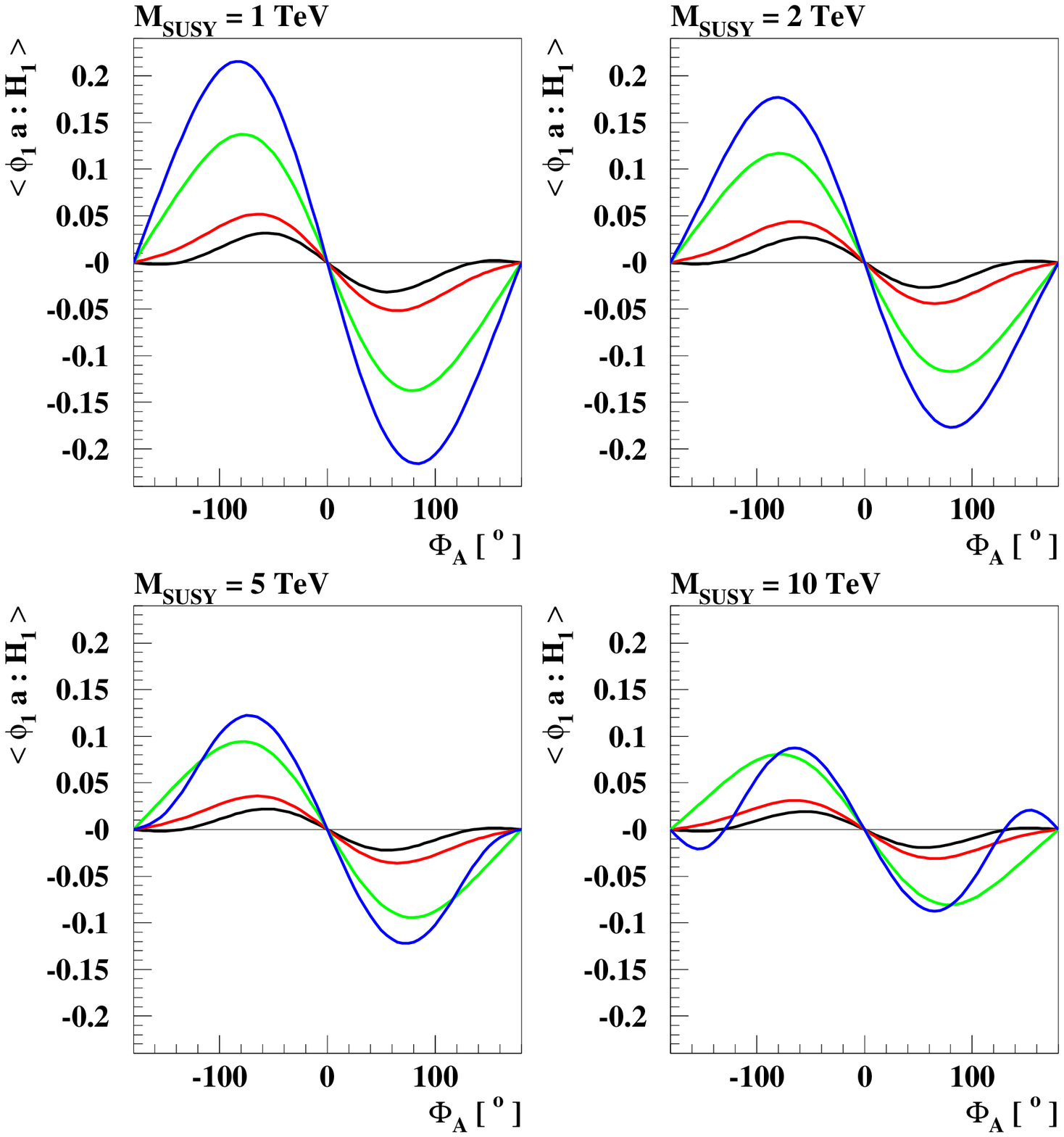}
\end{center}
\vspace{-1.0cm}
\caption{\it
The CP mixing quantity $\frac{2O_{\phi_1 1}O_{a 1}}{{O_{\phi_1 1}^2+O_{a 1}^2}}$
for the lightest mass eigenstate $H_1$ in scenarios with $M_{H^\pm}=500$ GeV
and $M_S=1$ TeV (upper left),
$M_S=2$ TeV (upper right),
$M_S=5$ TeV (lower left),
$M_S=10$ TeV (lower right).
The black, red, green and blue lines are for
$\tan\beta=5, 10, 30$ and $50$, respectively.
}
\label{fig:cpx.gs1gp.h1}
\end{figure}

We now consider predictions for two direct measures of CP violation in
the Higgs mass eigenstates $H_i$ (with $i =1,2,3$): 
\begin{equation}
\langle\phi_1 a : H_i\rangle \equiv 
\frac{2O_{\phi_1 i}O_{a i}}{O_{\phi_1 i}^2+O_{a i}^2}\,, \; \;
\langle\phi_2 a : H_i\rangle \equiv
\frac{2O_{\phi_2 i}O_{a i}}{O_{\phi_2 i}^2+O_{a i}^2}\, ,
\label{eq:gsgp}
\end{equation}
which characterize the mixtures between the CP-odd state $a$ and the
CP-even states $\phi_{1,2}$.  For instance, such CP-violating
expressions occur when studying CP violation in Higgs-boson decays to
fermions~\cite{Ellis:2004fs,Li:2015yla}. For a recent analysis of CP violation in
the decays $H_{1,2,3} \to \tau^+\tau^-$, see~\cite{Berge:2015nua}.
Figure~\ref{fig:cpx.gs1gp.h1} displays values of
$\langle\phi_1 a : H_1\rangle$ for the lightest neutral mass
eigenstate $H_1$, for the same scenarios $M_{H^\pm}=500$~GeV,
$M_S=1, 2, 5, 10$~TeV and $\tan\beta=5, 10, 30$ and $50$ discussed
previously.  We see that the values increase for smaller values of
$M_S$ and larger values of $\tan \beta$, and that values as large as
$\pm 0.22$ are possible for $\tan \beta = 50$ and $M_S = 1$~TeV.  Even
larger values would be possible for smaller values of $M_{H^\pm}$ and
$M_S$.

\begin{figure}[t!]
\vspace{-0.25cm}
\begin{center}
\includegraphics[width=9.5cm]{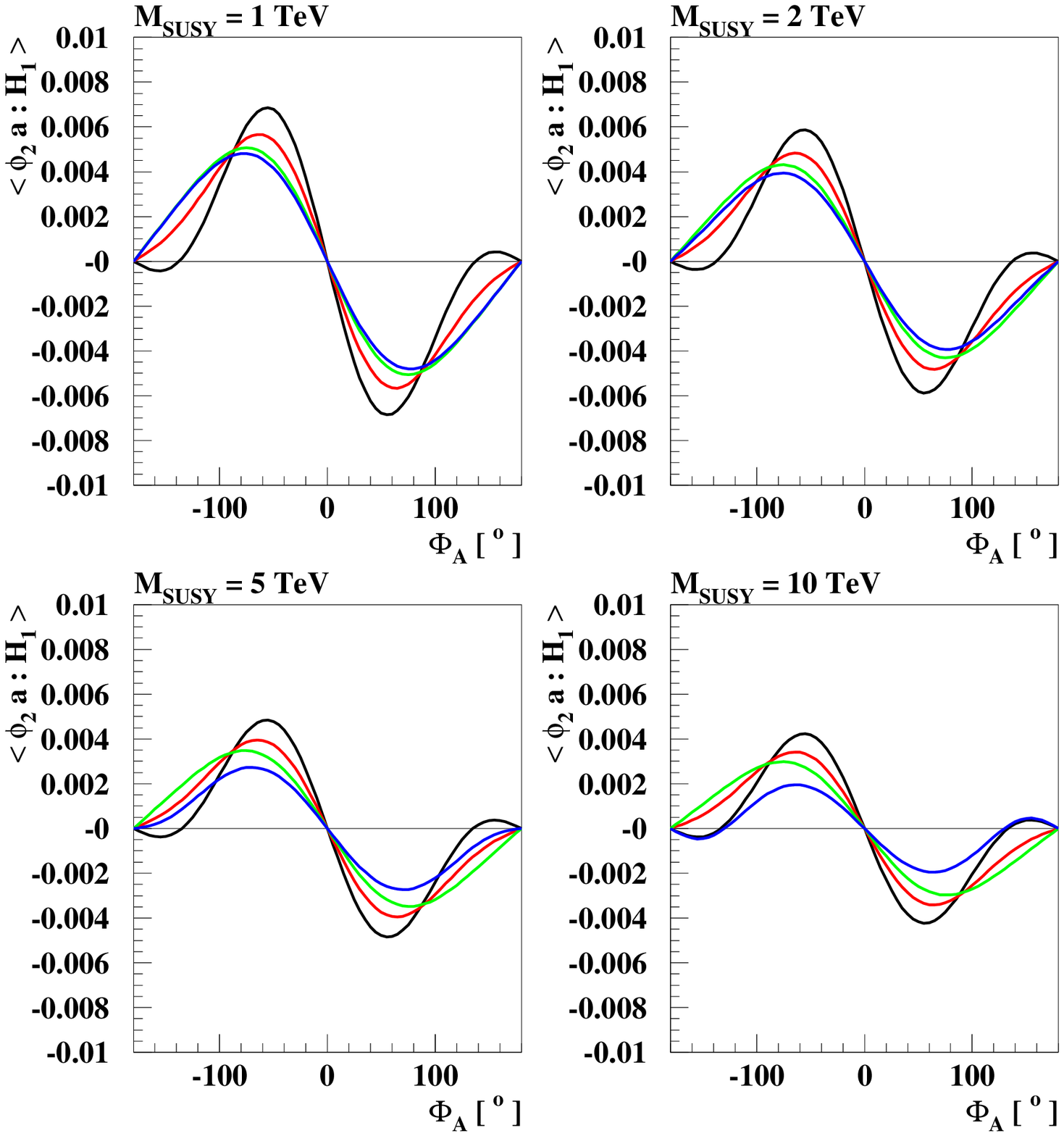}
\end{center}
\vspace{-0.5cm}
\caption{\it
As in Figure~\protect\ref{fig:cpx.gs1gp.h1}, but showing
the CP mixing quantity $\frac{2O_{\phi_2 1}O_{a 1}}{{O_{\phi_2 1}^2+O_{a 1}^2}}$
for the lightest mass eigenstate $H_1$.
}
\label{fig:cpx.gs2gp.h1}
\end{figure}

Fig~\ref{fig:cpx.gs2gp.h1} shows values of the other mixing
coefficient $\langle\phi_2 a : H_1 \rangle$ for $H_1$ in the same set
of CP-violating scenarios. This coefficient takes values $\lsim 0.007$
for $M_{H^\pm}=500$~GeV, $M_S = 1$~TeV and $\tan \beta = 5$,
decreasing for larger $M_{H^\pm}$, $M_S$ and $\tan \beta$.

\begin{figure}[t!]
\vspace{-0.75cm}
\begin{center}
\includegraphics[width=9.5cm]{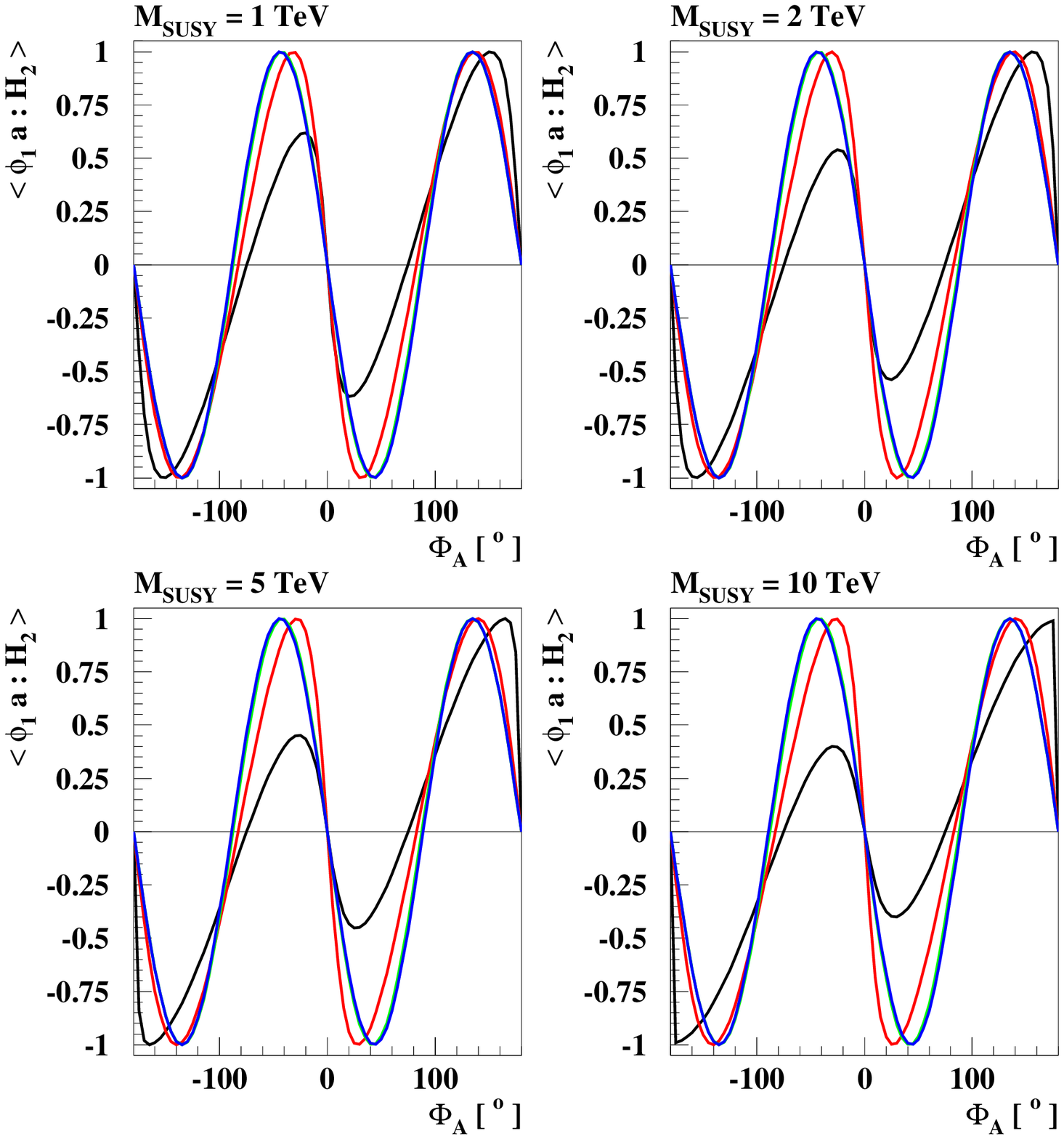}
\end{center}
\vspace{-0.75cm}
\caption{\it
As in Figure~\protect\ref{fig:cpx.gs1gp.h1}, but showing
the CP mixing quantity $\frac{2O_{\phi_1 2}O_{a 2}}{{O_{\phi_1 2}^2+O_{a 1}^2}}$
for the second mass eigenstate $H_2$.}
\label{fig:cpx.gs1gp.h2}
\end{figure}

\begin{figure}[t!]
\vspace{-0.5cm}
\begin{center}
\includegraphics[width=9.5cm]{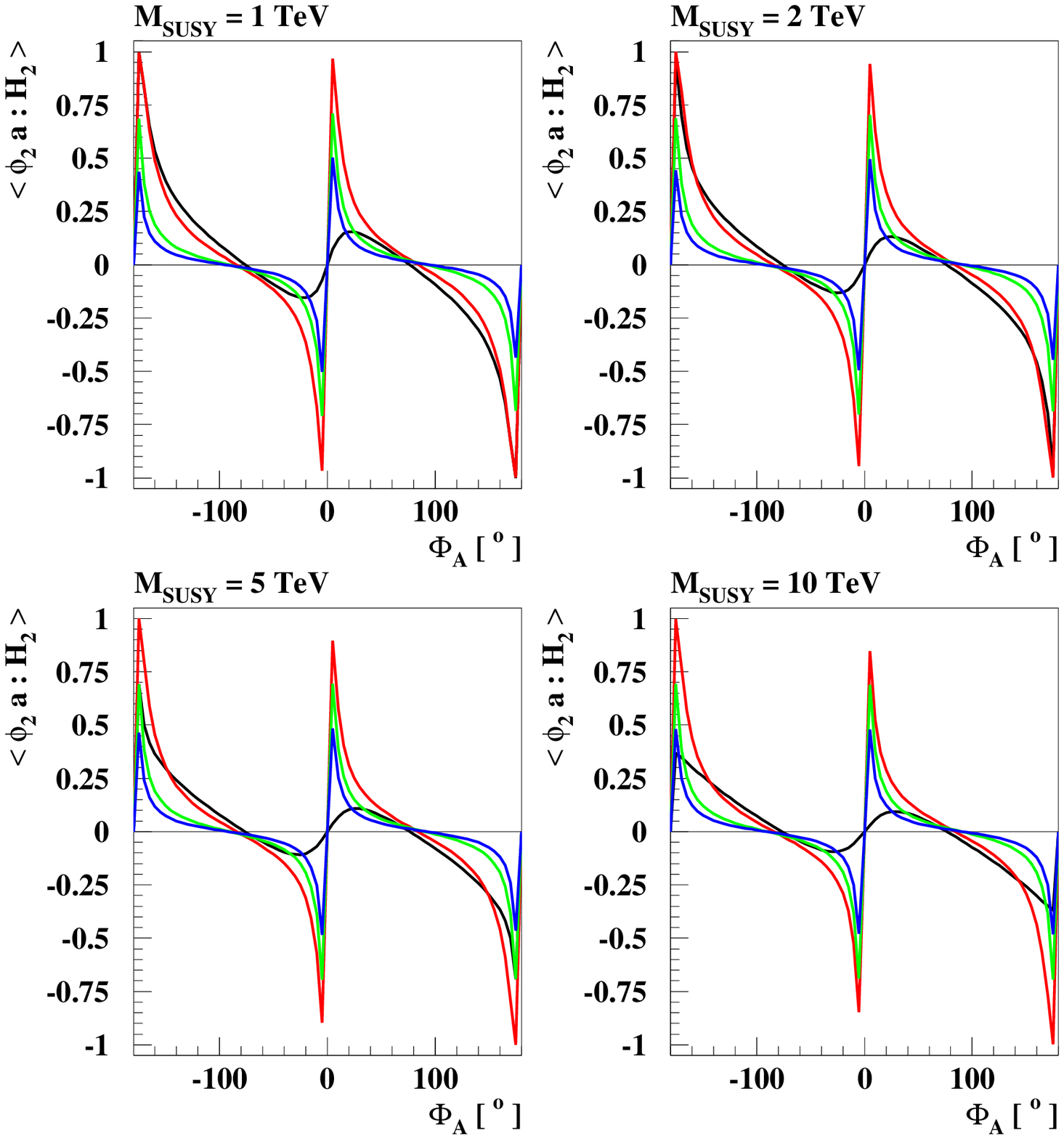}
\end{center}
\vspace{-0.75cm}
\caption{\it
As in Figure~\protect\ref{fig:cpx.gs1gp.h1}, but showing
the CP mixing quantity $\frac{2O_{\phi_2 2}O_{a 2}}{{O_{\phi_2 2}^2+O_{a 1}^2}}$
for the second mass eigenstate $H_2$.
}
\label{fig:cpx.gs2gp.h2}
\end{figure}

\begin{figure}[t!]
\vspace{-0.25cm}
\begin{center}
\includegraphics[width=9.5cm]{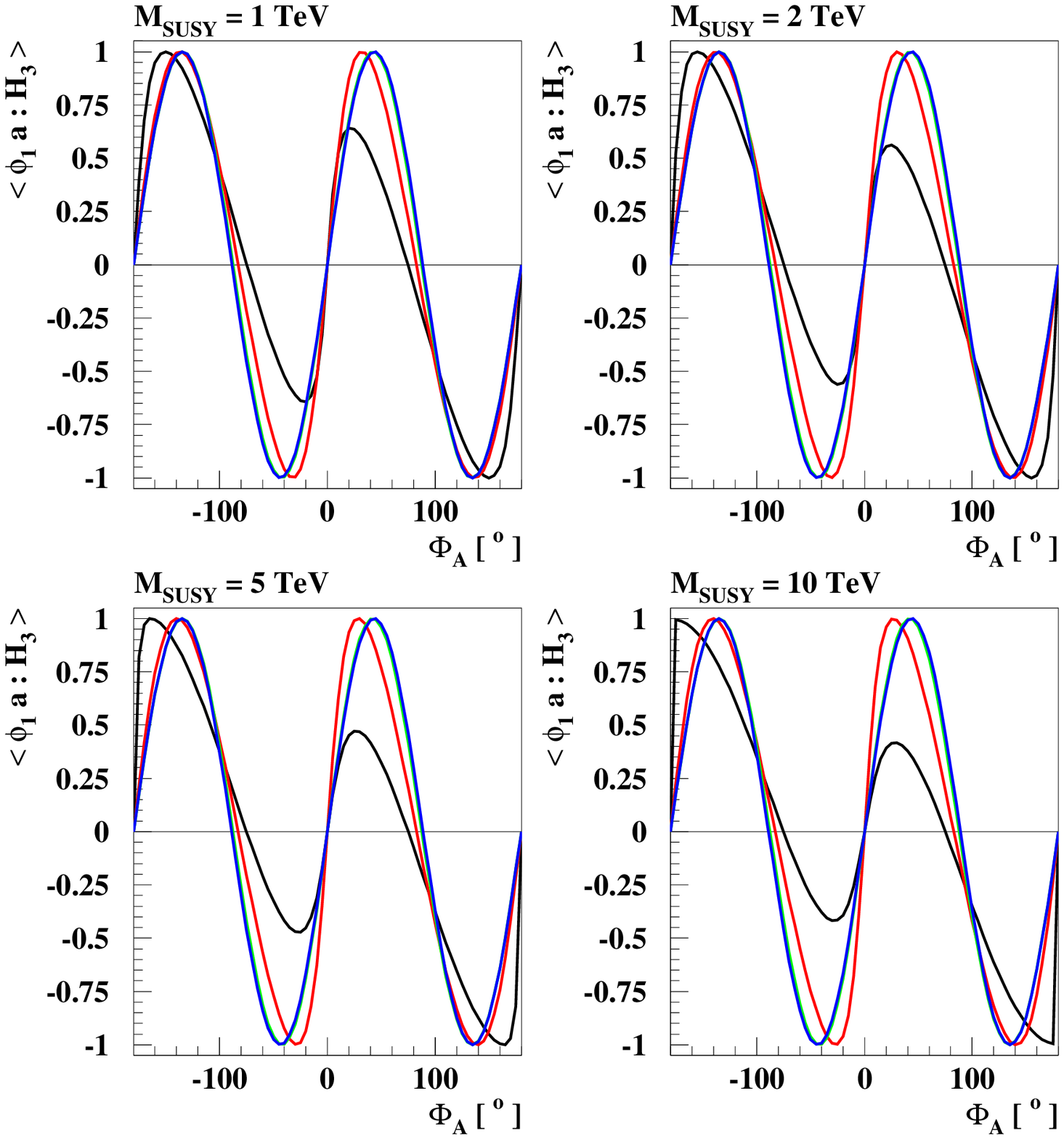}
\end{center}
\vspace{-0.75cm}
\caption{\it
As in Figure~\protect\ref{fig:cpx.gs1gp.h1}, but showing
the CP mixing quantity $\frac{2O_{\phi_1 3}O_{a 3}}{{O_{\phi_1 3}^2+O_{a 3}^2}}$
for the third mass eigenstate $H_3$.
}
\label{fig:cpx.gs1gp.h3}
\end{figure}

\begin{figure}[t!]
\vspace{-0.75cm}
\begin{center}
\includegraphics[width=9.5cm]{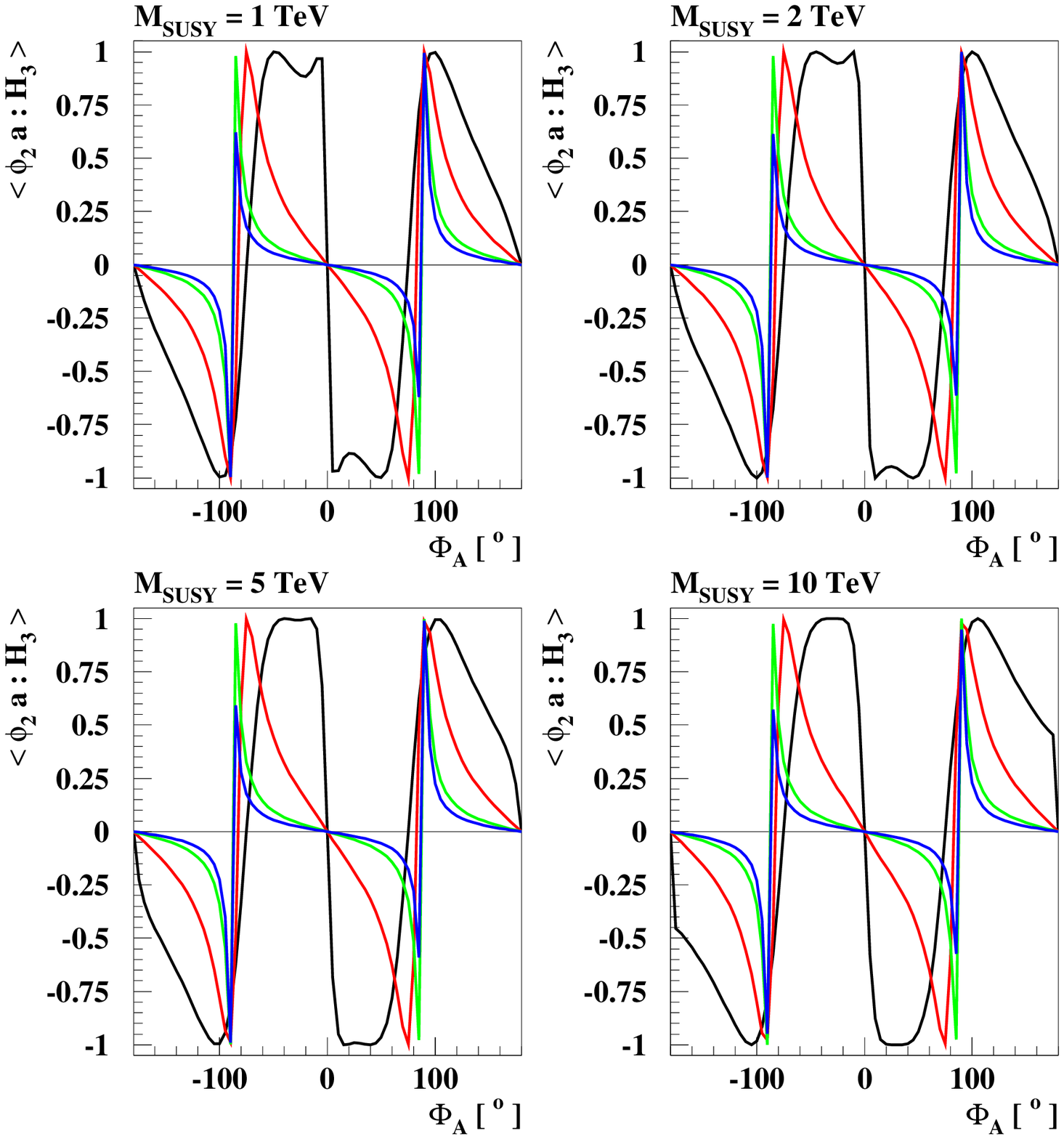}
\end{center}
\vspace{-0.75cm}
\caption{\it
As in Figure~\protect\ref{fig:cpx.gs1gp.h1}, but showing
the CP mixing quantity $\frac{2O_{\phi_2 3}O_{a 3}}{{O_{\phi_2 3}^2+O_{a 3}^2}}$
for the third mass eigenstate $H_3$.
}
\label{fig:cpx.gs2gp.h3}
\end{figure}

Similar results for the second mass eigenstate $H_2$ are shown in
Figs.~\ref{fig:cpx.gs1gp.h2} and \ref{fig:cpx.gs2gp.h2}, and for the
third mass eigenstate $H_3$ in Figs.~\ref{fig:cpx.gs1gp.h3} and
\ref{fig:cpx.gs2gp.h3}. We see here that the mixing quantities
(\ref{eq:gsgp}) can be {\it much larger} for the heavy mass
eigenstates $H_{2,3}$ than for the lightest mass eigenstate $H_1$,
attaining unity for many of the values of $M_S$ and $\tan \beta$
studied. This suggests, {\it a priori}, that the prospects for
observing CP violation would be enhanced for the heavier Higgs mass
eigenstates $H_2$ and $H_3$.

\begin{figure}[t!]
\vspace{-1.0cm}
\begin{center}
\includegraphics[width=16.5cm]{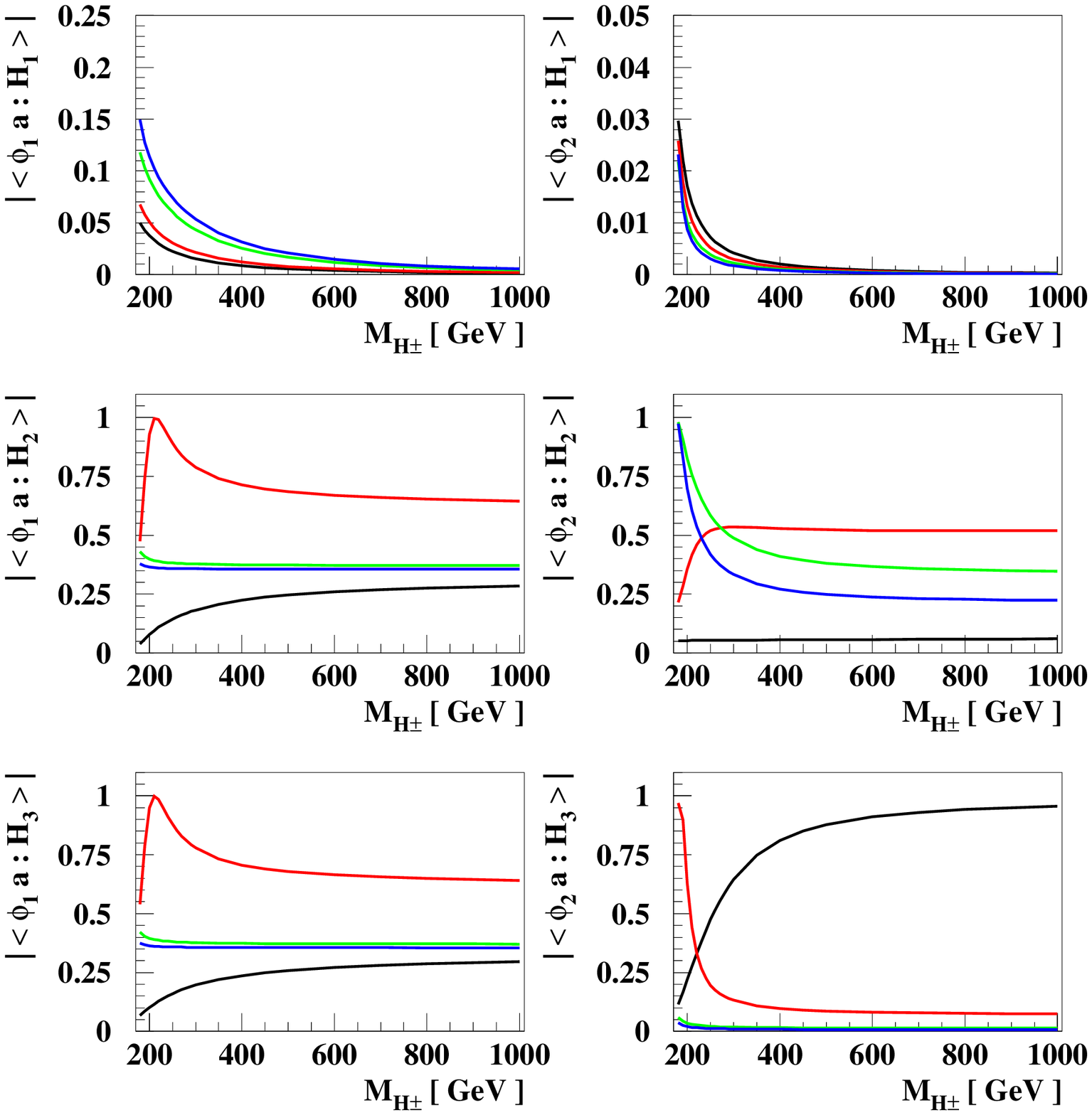}
\end{center}
\vspace{-0.5cm}
\caption{\it The magnitudes of the mixing quantities
  $\left|\langle\phi_1 a : H_i\rangle\right|$ (left) and
  $\left|\langle\phi_2 a : H_i\rangle\right|$ (right), defined in
  (\ref{eq:gsgp}), as functions of $M_{H^\pm}$ for $M_S=10$ TeV and
  $\Phi_A=10^\circ$.  The black, red, green and blue lines are for
  $\tan\beta=5, 10, 30$ and $50$, respectively.}
\label{fig:cpx.gsgp.mch.10000.10}
\end{figure}

\begin{figure}[t!]
\vspace{-1.0cm}
\begin{center}
\includegraphics[width=16.5cm]{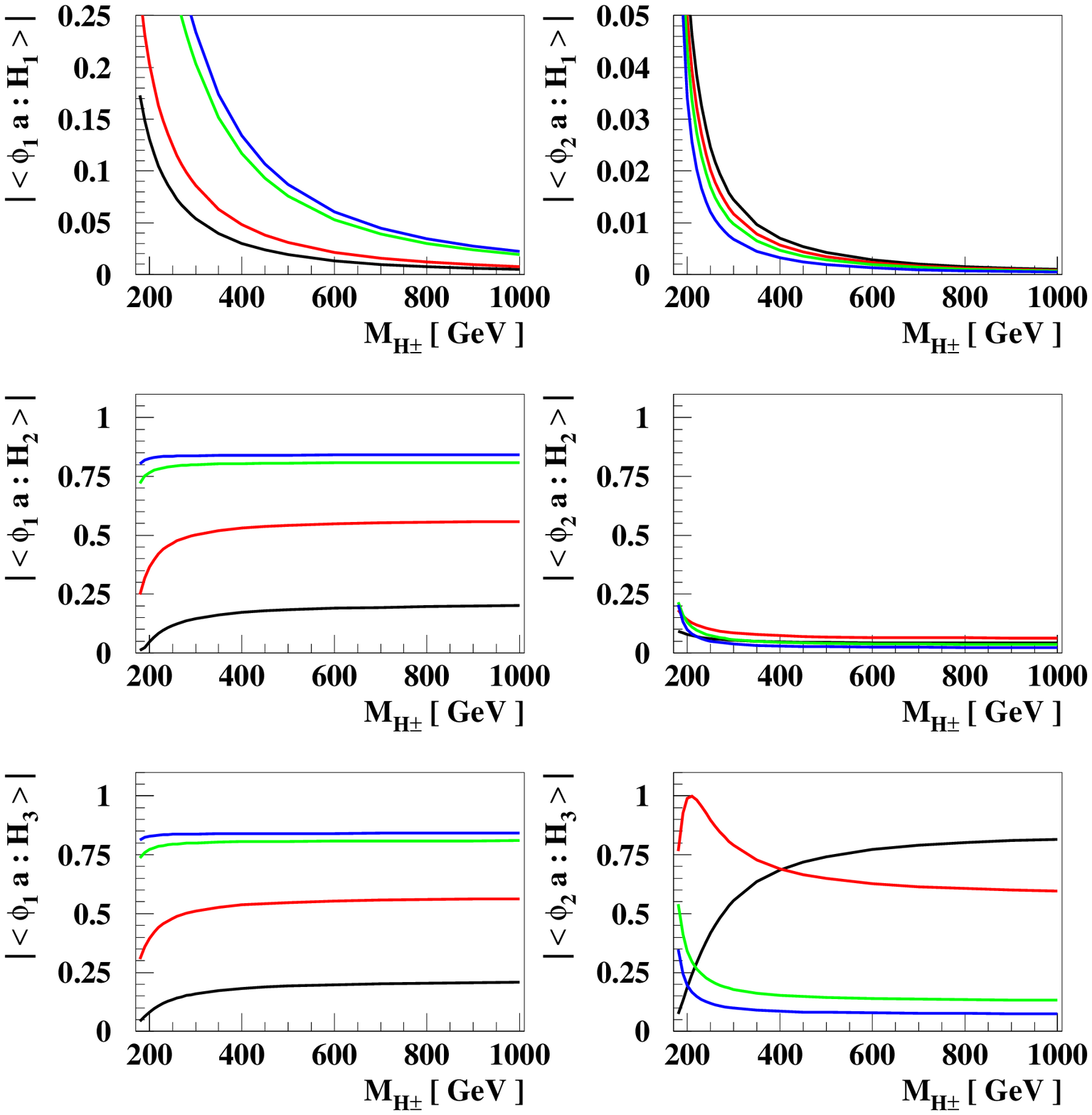}
\end{center}
\vspace{-0.5cm}
\caption{\it 
As in Figure~\ref{fig:cpx.gsgp.mch.10000.10} but for
for $M_S=10$ TeV and $\Phi_A=60^\circ$.
}
\label{fig:cpx.gsgp.mch.10000.60}
\end{figure}

These results illustrate the possible non-decoupling of CP-violating effects in
the  heavier  neutral  Higgs  bosons  $H_{2,3}$ for  large  values  of
$M_{H^\pm}$ and  $M_S$, whereas  the corresponding quantities  for the
lightest neutral Higgs mass eigenstate are expected to decouple.  This
is seen explicitly  in Figs.~\ref{fig:cpx.gsgp.mch.10000.10}, where we
display the  absolute values of $\langle\phi_1 a  : H_i\rangle$ (left)
and  $\langle\phi_2  a :  H_i\rangle$  (right)  for  $M_S=10$ TeV  and
$\Phi_A=10^\circ$ as  functions of $M_{H^\pm}$ for the  same values of
$\tan  \beta$ considered  previously.  The  corresponding  results for
$\Phi_A=60^\circ$ are shown in Figure~\ref{fig:cpx.gsgp.mch.10000.60}.
In the  case of $H_1$,  we see that  the mixing quantities $\to  0$ at
large $M_{H^\pm}$,  as expected, whereas in  general the corresponding
coefficients for the heavy neutral Higgs mass eigenstates $H_{2,3}$ do
not vanish, and  retain large values even for  very large $M_{H^\pm}$.
Thus, large CP-violating effects in the couplings of these states are a
robust signature of the CP-violating scenarios discussed in this paper.

\section{Conclusions}

We present  new MSSM scenarios with explicit  CP violation that
contain heavy  Higgs bosons in the few to several hundred GeV range 
and  are consistent with  constraints from
Run~I of the LHC.  The scenarios  suggested here are similar in spirit
to the  CPX scenarios   previously proposed , and  have phenomenological
implications that can be tested during  Run II of the LHC. In light of
this, we call them \textcolor{red}{CP}\textcolor{green}{X}{\it
  4}\textcolor{blue}{LHC}  benchmark scenarios.

In this work we explicitly demonstrate that, although CP violation and
other new-physics effects decouple from the lightest Higgs boson
sector for sufficiently large charged Higgs boson masses and soft
SUSY-breaking scales $M_S$, they can still be significant in the MSSM
heavy Higgs boson sector.  Large masses of the supersymmetric
particles also help to maintain agreement with limits on EDMs and
other low-energy observables.  We consider scenarios in which the
charged Higgs bosons $H^\pm$ and the two heavier neutral Higgs
bosons~$H_{2,3}$ could be much lighter than all third generation
supersymmetric scalar fermions, which are assumed to have masses
$M_S \gsim 2$~TeV.  In light of this possibility, we have revisited
previous calculations by considering improved matching and
renormalization group (RG) effects, specifically including two-loop RG
effects in the two-Higgs-doublet model (2HDM) that is effective
between the heavy Higgs scale~$M_{H^\pm}$ and the SUSY scale~$M_S$.

We compare our new results 
with  those obtained with the previous  code version {\tt  CPsuperH2.3}. 
We also discuss the  specific \textcolor{red}{CP}\textcolor{green}{X}{\it
  4}\textcolor{blue}{LHC}  benchmark scenarios relevant for the analysis of
Higgs physics at the LHC,
with particular emphasis on the masses of the
heavier  neutral Higgs  bosons $H_{2,3}$  and on  the CP-violating effects
they may  manifest. These offer interesting prospects  for future runs
of the LHC and future colliders.

All the improvements discussed in this study are being incorporated in
a  new  version  of  the  public  code  {\tt  CPsuperH},  called  {\tt
  CPsuperH3.0}.  The  numerical results  presented here  have been
obtained by means of a  preliminary $\beta$-version of this code.  The
detailed features of  {\tt CPsuperH3.0} will be fully  described in an
upcoming release note~\footnote{The  current $\beta$-version is available upon
request to J.~S.~Lee.}.

\subsection*{Acknowledgements}
Fermilab is operated by Fermi Research Alliance, LLC under Contract
No. DE-AC02-07CH11359 with the U.S. Department of Energy, and
University of Chicago is supported in part by U.S. Department of
Energy grant number DE-FG02-13ER41958.  The work of J.E. is supported
in part by the London Centre for Terauniverse Studies (LCTS), using
funding from the European Research Council via the Advanced
Investigator Grant 267352, and in part by STFC (UK) via the research
grant ST/L000326/1. The work of J.S.L. is supported by the National
Research Foundation of Korea (NRF) grant (No.  2013R1A2A2A01015406).
The work of A.P. is supported by the Lancaster-Manchester-Sheffield
Consortium for Fundamental Physics under STFC grant ST/L000520/1.
Work at ANL is supported in part by the U.S. Department of Energy
under Contract No. DE-AC02-06CH11357.  M.C. and C.W.  thank for its
hospitality the Aspen Center for Physics, which is supported by the
National Science Foundation under Grant No. PHYS-1066293.

\newpage
\section*{Appendices}

In the appendices that follow, we present the relevant Renormalization
Group Equations (RGEs) that  are applicable  above the  three typical
thresholds: (i)  the top-quark mass  $m_t$, (ii) the heavy  Higgs mass
$M_H\equiv M_{H^+}$, and (iii) the soft SUSY-breaking scale $M_S$.

It is convenient to write the RGEs in the form
$$
\frac{{\rm d}c}{{\rm d}t}\ =\ \sum_{n=1}\kappa^n\,\beta_c^{(n)}\;,
$$
where $c$ stands for any  kinematic parameter, such as~quartic, gauge and
Yukawa couplings, anomalous dimensions, and $\tan\beta$. In addition,
we use the abbreviations: $t\equiv\ln Q$ and $\kappa=1/(4\pi)^2$.

\def\theequation{\Alph{section}.\arabic{equation}}
\begin{appendix}
\setcounter{equation}{0}
\section{SM RGEs}\label{RGE:SM}

For scales of $Q$,  for which $M_{H_1} \sim m_t < Q  < M_H$, one needs
to consider  the RGEs of the  SM. To~properly take  into consideration
intermediate  particle threshold  effects,  we introduce the  short-hand
notation for the step function
$$
\Theta_X\ \equiv\ \left\{
\begin{array}{cl}
1 & {\rm for}~~ Q\geq M_X\,, \\
0 & {\rm for}~~ Q < M_X\, .
\end{array} \right.
$$

Upon neglecting the  Yukawa couplings of the first  two generations of
quarks    and    leptons,   the    one-loop    SM    RGEs   that    we
use~\cite{Draper:2013oza,Arason:1991ic} read:
\begin{eqnarray}
\beta^{(1)}_\lambda &=& 
12 \lambda^2 + 4 \lambda ( 3y_t^2 + 3y_b^2 + y_\tau^2) - 4 ( 3y_t^4 + 3y_b^4 +
y_\tau^4 ) 
\nonumber \\ &&
- 9 \lambda \left( g^2 + \frac{1}{3} g'^2 \right) 
+ \frac{9}{4} \left( g^4 + \frac{2}{3} g'^2 g^2 + \frac{1}{3} g'^4 \right)
\nonumber \\ &&
+\left(6\lambda g^2-5g^4+8g^4s_\beta^2c_\beta^2\right)
\Theta_{\widetilde{H}}\Theta_{\widetilde{W}}
\nonumber \\ &&
-2g^2g'^2\Theta_{\widetilde{H}}\Theta_{\widetilde{W}}\Theta_{\widetilde{B}}
+\left(2\lambda g'^2-g'^4\right)
\Theta_{\widetilde{H}}\Theta_{\widetilde{B}} \, ,
\\[2mm]
\beta_\lambda^{(2)} &=&
-78 \lambda^3 - 72 \lambda^2 y_t^2 + 80 \lambda g_s^2 y_t^2 - 3 \lambda y_t^4 
- 64g_s^2 y_t^4  + 60 y_t^6 \, ,
\\[2mm]
\beta_\lambda^{(3)} &=& 
\frac{\lambda^3}{2} \left( 6011.35 \frac{\lambda}{2} + 873 y_t^2 \right) + 
\lambda^2 y_t^2 (1768.26 y_t^2 + 160.77 g_s^2)
\nonumber \\ &&
 + 2 \lambda y_t^2 (-223.382 y_t^4 - 662.866 g_s^2 y_t^2 + 356.968 g_s^4)
\nonumber \\ &&
 + 4y_t^4 (-243.149 y_t^4 + 250.494 g_s^2 y_t^2 - 50.201 g_s^4) \, ,
\end{eqnarray}
\begin{eqnarray}
\beta_{g_s}^{(1)} &=& -g_s^3 \left[ 11 - \frac{2}{3} N_f \right] \, ,
\\[2mm]
\beta_{g_s}^{(2)} &=& -g_s^3 \left[ 
\left( 102 - \frac{38}{3} N_f \right) g_s^2 
-\frac{3}{4}N_fg^2-\frac{11}{36}N_fg'^2 
+2 y_t^2+2 y_b^2
\right] \, ,
\end{eqnarray}
\begin{eqnarray}
\beta_{y_t}^{(1)} &=& y_t \left[ 
\frac{9}{2} y_t^2 + \frac{3}{2} y_b^2 + y_\tau^2 - 8 g_s^2 -
\frac{9}{4} g^2 - \frac{17}{12} g'^2 \right]
\nonumber \\
&& 
+\frac{3}{2}y_t g^2 \Theta_{\widetilde{H}}\Theta_{\widetilde{W}}
+\frac{1}{2}y_t g'^2 \Theta_{\widetilde{H}}\Theta_{\widetilde{B}} \, ,
\\[2mm]
\beta_{y_t}^{(2)} &=& y_t \left[ 
\frac{3}{2} \lambda^2 - 6\lambda y_t^2 - \left( \frac{404}{3} -
\frac{40}{9} N_f \right) g_s^4 + 36 g_s^2 y_t^2 - 12y_t^4 \right] \, ,
\\[2mm]
\beta_{y_t}^{(3)} &=& y_t \left[  -\frac{9}{2} \lambda^3 
+\frac{15}{16} \lambda^2 y_t^2 + \lambda y_t^2
(99 y_t^2 + 8 g_s^2)\right.
\nonumber \\ && \left.  \vphantom{\frac{15}{16}} \hspace{0.5cm}
 + 58.6028 y_t^6 - 157 y_t^4 g_s^2 + 363.764 y_t^2 g_s^4 - 619.35 g_s^6 \right] \, ,
\end{eqnarray}
\begin{eqnarray}
\beta_{y_b}^{(1)} = y_b \left[ 
\frac{3}{2} y_t^2 + \frac{9}{2} y_b^2 + y_\tau^2 - 8 g_s^2 -
\frac{9}{4} g^2 - \frac{5}{12} g'^2 \right] \, ,
\end{eqnarray}
\begin{eqnarray}
\beta_{y_\tau}^{(1)} = y_\tau \left[ 
3 y_t^2 + 3 y_b^2 + \frac{5}{2}y_\tau^2  -
\frac{9}{4} g^2 - \frac{45}{12} g'^2 \right] \, ,
\end{eqnarray}
\begin{eqnarray}
\beta_{g'}^{(1)} &= &
\left(\frac{2}{3}N_f+\frac{1}{10}\right)\frac{5}{3}\,g'^3 \, ,
\nonumber \\[2mm]
\beta_{g'}^{(2)} &= &
\left(\frac{22}{9}N_fg_s^2-\frac{17}{6}y_t^2\right)g'^3 \, ,
\nonumber \\[2mm]
\beta_{g}^{(1)} &= &
\left(-\frac{22}{3}+\frac{2}{3}N_f+\frac{1}{6}\right)g^3 \, ,
\nonumber \\[2mm]
\beta_{g}^{(2)} &= &
\left(2N_fg_s^2-\frac{3}{2}y_t^2\right)g^3 \, ,
\end{eqnarray}
\begin{eqnarray}
  \label{gammaSM}
\gamma^{(1)} = 
\frac{9}{4} \left(g^2 + \frac{1}{3} g'^2 \right)
-(3 y_t^2 + 3 y_b^2 + y_\tau^2  ) \, .
\end{eqnarray}

\setcounter{equation}{0}
\section{One-Loop 2HDM RGEs}\label{RGE:2HDM1loop}

For  RG scales  $Q$  between  $M_H$ and  $M_S$,  the effective  theory
becomes a general  2HDM, whilst for $Q> M_S$ the  theory becomes fully
supersymmetric and the quartic couplings $\lambda_{5,6,7}$ do not run.
Here, we give the RGEs of the  general 2HDM at the one-loop level, and
relegate  to  Appendix~\ref{RGE:2HDM2loop}  the  presentation  of  the
two-loop results.

As  before,  we  neglect  the   Yukawa  couplings  of  the  first  two
generations of  quarks and leptons, and  note that in addition  to the
change  of  normalizations   given  in~(\ref{eq:match.hh.1}),  we  use
$t    =    \ln(Q)$    instead     of    $\ln(Q^2)$    as    used    in
Ref.~\cite{Haber:1993an}.      Thus,     adapting      the     results
of~\cite{Haber:1993an}, the one-loop 2HDM RGEs may be listed as
follows:
\begin{eqnarray}
\beta_{\lambda_{1}}^{(1)} &=& -\left\{24\lambda^2_{1}+\lambda_3^2
+\left(\lambda_3+\lambda_4\right)^2
+4\lambda_5^2+12\lambda_{6}^2+\frac{3}{8}
[2g^4+(g^2+g'^2)^2]\right\}\Theta_Z
\nonumber \\ &&
- N_C\Bigg\{-2h_b^4 \Theta_Z
 +\left(h_b^2-\frac{1}{4} g'^2Y_D\right)^2\Theta_{\widetilde D_3}
 +\left(\frac{1}{4} g'^2Y_U\right)^2\Theta_{\widetilde U_3}
\nonumber \\ && \hspace{1.0cm} 
+\left[h_b^4-\frac{1}{2} h_b^2(g'^2Y_Q+g^2)
+\frac{1}{8}(g^4+g'^4Y_Q^2)\right]\Theta_{\widetilde Q_3}\Bigg\}
\nonumber \\ &&
- N_C\sum_{i=1}^2\Bigg\{
  \left(\frac{1}{4} g'^2Y_D\right)^2\Theta_{\widetilde D_i}
 +\left(\frac{1}{4} g'^2Y_U\right)^2\Theta_{\widetilde U_i}
 +\frac{1}{8}(g^4+g'^4Y_Q^2)\Theta_{\widetilde Q_i}\Bigg\}
\nonumber \\ &&
-\Bigg\{-2h_\tau^4 \Theta_Z
 +\left(h_\tau^2-\frac{1}{4} g'^2Y_E\right)^2\Theta_{\widetilde E_3}
+\left[h_\tau^4-\frac{1}{2} h_\tau^2(g'^2Y_L+g^2)
+\frac{1}{8}(g^4+g'^4Y_L^2)\right]\Theta_{\widetilde L_3}\Bigg\}
\nonumber \\ &&
- \sum_{i=1}^2\Bigg\{
  \left(\frac{1}{4} g'^2Y_E\right)^2\Theta_{\widetilde E_i}
 +\frac{1}{8}(g^4+g'^4Y_L^2)\Theta_{\widetilde L_i}\Bigg\}
\nonumber \\ &&
+\frac{5}{2} g^4\Theta_{\widetilde H}\Theta_{\widetilde W}
+g'^2g^2\Theta_{\widetilde H}\Theta_{\widetilde W}\Theta_{\widetilde B}
+\frac{1}{2} g'^4\Theta_{\widetilde H}\Theta_{\widetilde B}
-4\lambda_{1}\gamma_{1}^{(1)}\,,
\end{eqnarray}
where $N_C=3$, $Y_Q=1/3$, $Y_U=-4/3$, $Y_D=2/3$, $Y_L=-1$, $Y_E=2$, and
\begin{eqnarray}
\beta_{\lambda_{2}}^{(1)} &=& -\left\{24\lambda^2_{2}+\lambda_3^2
+\left(\lambda_3+\lambda_4\right)^2
+4\lambda_5^2+12\lambda_{6}^2+\frac{3}{8}
[2g^4+(g^2+g'^2)^2]\right\}\Theta_Z
\nonumber \\ &&
- N_C\Bigg\{-2h_t^4 \Theta_t\Theta_Z
 +\left(\frac{1}{4} g'^2Y_D\right)^2\Theta_{\widetilde D_3}
 +\left(h_t^2+\frac{1}{4} g'^2Y_U\right)^2\Theta_{\widetilde U_3}
\nonumber \\ && \hspace{1.0cm}
+\left[h_t^4+\frac{1}{2} h_t^2(g'^2Y_Q-g^2)
+\frac{1}{8}(g^4+g'^4Y_Q^2)\right]\Theta_{\widetilde Q_3}\Bigg\}
\nonumber \\ &&
- N_C\sum_{i=1}^2\Bigg\{
  \left(\frac{1}{4} g'^2Y_D\right)^2\Theta_{\widetilde D_i}
 +\left(\frac{1}{4} g'^2Y_U\right)^2\Theta_{\widetilde U_i}
 +\frac{1}{8}(g^4+g'^4Y_Q^2)\Theta_{\widetilde Q_i}\Bigg\}
\nonumber \\ && 
- \sum_{i=1}^3\Bigg\{
 \left(\frac{1}{4} g'^2Y_E\right)^2\Theta_{\widetilde E_i}
+ \frac{1}{8}(g^4+g'^4Y_L^2)\Theta_{\widetilde L_i}\Bigg\}
\nonumber \\ && 
+\frac{5}{2} g^4\Theta_{\widetilde H}\Theta_{\widetilde W}
+g'^2g^2\Theta_{\widetilde H}\Theta_{\widetilde W}\Theta_{\widetilde B}
+\frac{1}{2} g'^4\Theta_{\widetilde H}\Theta_{\widetilde B}
-4\lambda_{2}\gamma_{2}^{(1)}\, ,
\end{eqnarray}
\begin{eqnarray}
\beta_{\lambda_3}^{(1)}&=&-2\Bigg\{2(\lambda_1+\lambda_2)(3\lambda_3+\lambda_4)
+2\lambda_3^2+\lambda_4^2+4\lambda_5^2+2\lambda_6^2+2\lambda_7^2
+8\lambda_6\lambda_7
\nonumber \\ &&  \hspace{0.8cm}
+\frac{3}{8}
\big[2g^4+(g^2-g'^2)^2\big]\Bigg\}\Theta_Z
\nonumber \\ &&
-2 N_C\Bigg\{-2h_t^2 h_b^2 \Theta_{t}\Theta_Z
+\frac{1}{4} g'^2Y_D\left(h_b^2-\frac{1}{4} g'^2Y_D\right)\Theta_{\widetilde D_3}
\nonumber \\ && \hspace{1.5cm}
-\frac{1}{4} g'^2Y_U\left(h_t^2+\frac{1}{4} g'^2Y_U\right)\Theta_{\widetilde U_3}
+h_t^2h_b^2\Theta_{\widetilde U_3}\Theta_{\widetilde D_3}
\nonumber \\ && \hspace{1.5cm}
+\left[h_t^2h_b^2-\frac{1}{4} h_t^2(g'^2Y_Q+g^2) 
+\frac{1}{4} h_b^2(g'^2Y_Q-g^2)+\frac{1}{8}(g^4-g'^4Y_Q^2)\right]
\Theta_{\widetilde Q_3}\Bigg\}
\nonumber \\ &&
-2 N_C\sum_{i=1}^2\Bigg\{
 -\left(\frac{1}{4} g'^2Y_D\right)^2\Theta_{\widetilde D_i}
 -\left(\frac{1}{4} g'^2Y_U\right)^2\Theta_{\widetilde U_i}
 +\frac{1}{8}(g^4-g'^4Y_Q^2)\Theta_{\widetilde Q_i}\Bigg\}
\nonumber \\ &&
-2 \Bigg\{
\frac{1}{4} g'^2Y_E\left(h_\tau^2-\frac{1}{4} g'^2Y_E\right)\Theta_{\widetilde E_3}
+\left[
\frac{1}{4} h_\tau^2(g'^2Y_L-g^2)+\frac{1}{8}(g^4-g'^4Y_L^2)\right]
\Theta_{\widetilde L_3}\Bigg\}
\nonumber \\ &&
-2 \sum_{i=1}^2\Bigg\{
 -\left(\frac{1}{4} g'^2Y_E\right)^2\Theta_{\widetilde E_i}
 +\frac{1}{8}(g^4-g'^4Y_L^2)\Theta_{\widetilde L_i}\Bigg\}
\nonumber \\ &&
+5 g^4\Theta_{\widetilde W}\Theta_{\widetilde H}
-2g'^2g^2\Theta_{\widetilde W}\Theta_{\widetilde B}\Theta_{\widetilde H} 
+ g'^4\Theta_{\widetilde B}\Theta_{\widetilde H} 
-2\lambda_3(\gamma_1^{(1)}+\gamma_2^{(1)})\, ,
\end{eqnarray}
\begin{eqnarray}
\beta_{\lambda_4}^{(1)}&=&
-2\left[ \lambda_4 (2\lambda_1 +2\lambda_2+4\lambda_3+2\lambda_4)
+ 16\lambda_5^2+5\lambda_6^2+5\lambda_7^2+2\lambda_6\lambda_7
+\frac{3}{2} g^2g'^2\right]\Theta_Z
\nonumber \\ &&
-2 N_C\Bigg\{ 2h_t^2h_b^2\Theta_{t}\Theta_Z
-h_t^2h_b^2 \Theta_{\widetilde U_3}\Theta_{\widetilde D_3}
-\left(h_t^2-\frac{1}{2} g^2\right)\left(h_b^2-\frac{1}{2} g^2\right) 
\Theta_{\widetilde Q_3} \Bigg\}
\nonumber \\ &&
+2 N_C\sum_{i=1}^2 \left(\frac{1}{2} g^2\right)^2\Theta_{\widetilde Q_i}
-g^2\left(h_\tau^2-\frac{1}{2} g^2\right) \Theta_{\widetilde L_3}
+\sum_{i=1}^2 \frac{1}{2} g^4\Theta_{\widetilde L_i}
\nonumber \\ &&
-4g^4\Theta_{\widetilde W}\Theta_{\widetilde H}
+4g'^2g^2\Theta_{\widetilde W}\Theta_{\widetilde B}\Theta_{\widetilde H}
-2\lambda_4(\gamma_1^{(1)}+\gamma_2^{(1)})\, ,
\end{eqnarray}
\begin{eqnarray}
\beta_{\lambda_5}^{(1)} &=&
-\left[  2\lambda_5 (2\lambda_1+2\lambda_2+4\lambda_3+6\lambda_4)
+5\left(\lambda_6^2+\lambda_7^2\right)
+2\lambda_6\lambda_7\right]\Theta_Z
-2\lambda_5(\gamma_1^{(1)}+\gamma_2^{(1)})\, ,
\nonumber \\
\end{eqnarray}
\begin{eqnarray}
\beta_{\lambda_6}^{(1)} &=&
-2\left[ \lambda_6\left(12\lambda_1+3\lambda_3
+4\lambda_4+10\lambda_5\right) +\lambda_7 \left(
3\lambda_3+2\lambda_4+2\lambda_5\right)\right]\Theta_Z
-\lambda_6(3\gamma_1^{(1)}+\gamma_2^{(1)})\, ,
\nonumber \\
\end{eqnarray}
\begin{eqnarray}
\beta_{\lambda_7}^{(1)} &=&
-2\left[ \lambda_7\left(12\lambda_2+3\lambda_3
+4\lambda_4+10\lambda_5\right) +\lambda_6 \left(
3\lambda_3+2\lambda_4+2\lambda_5\right)\right]\Theta_Z
-\lambda_7(\gamma_1^{(1)}+3\gamma_2^{(1)})\, ,
\nonumber \\
\end{eqnarray}
\begin{eqnarray}
\gamma_1^{(1)}&=&
{1\over 4}
\left[\left(9g^2+3g'^2-4
(N_Ch_b^2+h_\tau^2)\right)\Theta_Z
-6g^2\Theta_{\widetilde W}\Theta_{\widetilde H}
-2g'^2\Theta_{\widetilde B}\Theta_{\widetilde H}\right]\,,
\\[2mm]
\gamma_2^{(1)}&=&
{1\over 4}
\left[\left(9g^2+3g'^2-4N_Ch_t^2\Theta_t\right)\Theta_Z
-6g^2\Theta_{\widetilde W}\Theta_{\widetilde H}
-2g'^2\Theta_{\widetilde B}\Theta_{\widetilde H}\right]\, ,
\hspace{2.0cm}
\end{eqnarray}
\begin{eqnarray}
\beta_{h_t}^{(1)}&=&\left(
\frac{9}{2} h_t^2+\frac{1}{2} h_b^2-8g_s^2
-\frac{9}{4} g^2-\frac{17}{12} g'^2\right)h_t \, , \\
\beta_{h_b}^{(1)}&=&\left(
\frac{9}{2} h_b^2+\frac{1}{2} h_t^2+h_\tau^2-8g_s^2
-\frac{9}{4} g^2-\frac{5}{12} g'^2\right)h_b \, , \\
\beta_{h_\tau}^{(1)}&=&\left(
\frac{5}{2} h_\tau^2+3h_b^2
-\frac{9}{4} g^2-\frac{15}{4} g'^2\right)h_\tau \, ,
\end{eqnarray}
\begin{eqnarray}
\beta_{g'}^{(1)}&=&\Bigg\{
\frac{1}{4}N_C\sum_{i=1}^3\left[
2Y_Q^2\left(2\Theta_t+\Theta_{\widetilde Q_i}\right)
+Y_U^2\left(2\Theta_t+\Theta_{\widetilde U_i}\right)
+Y_D^2\left(2\Theta_t+\Theta_{\widetilde D_i}\right)\right]
\nonumber \\ &&
+\frac{1}{4}\sum_{i=1}^3\left[
2Y_L^2\left(2\Theta_t+\Theta_{\widetilde L_i}\right)
+Y_E^2\left(2\Theta_t+\Theta_{\widetilde E_i}\right)\right]
+\frac{1}{2}N_H\Theta_t+N_{\widetilde H}\Theta_{\widetilde H}
\Bigg\}\,\frac{g'^3}{3}\ ,
\nonumber \\ 
\end{eqnarray}
with $N_H=2$ and  $N_{\widetilde H}=2$,
\begin{eqnarray}
\beta_{g}^{(1)}&=&\Bigg\{
\frac{1}{2} N_C \sum_{i=1}^3\left(2\Theta_t+\Theta_{\widetilde Q_i}\right)
+\frac{1}{2} \sum_{i=1}^3\left(2\Theta_t+\Theta_{\widetilde L_i}\right)
+\frac{1}{2} N_H\Theta_t+N_{\tilde H}\Theta_{\widetilde H}
+4N_{\tilde W}\Theta_{\widetilde W}-22\Theta_t
\Bigg\}\,\frac{g^3}{3}\ ,
\nonumber \\ 
\end{eqnarray}
with $N_{\tilde W}=1$,
\begin{eqnarray}
\beta_{g_s}^{(1)}&=&\Bigg\{2N_{f}+
\sum_{i=1}^3\left(
 \Theta_{\widetilde Q_i}
+\frac{1}{2}\Theta_{\widetilde U_i}
+\frac{1}{2}\Theta_{\widetilde D_i}\right)
+6N_{\tilde g}\Theta_{\tilde g}-33
\Bigg\}\,\frac{g_s^3}{3}\ ,
\nonumber \\ 
\end{eqnarray}
with 
$N_f=\Theta_t+\Theta_b+\Theta_c+\Theta_s+\Theta_d+\Theta_u$ and
$N_{\tilde g}=1$,
\begin{eqnarray}
\beta_{\tan\beta}^{(1)}&=&\left(\gamma_2^{(1)}-\gamma_1^{(1)}\right)\tan\beta
\nonumber \\
&=&\left(-N_Ch_t^2\Theta_t\Theta_Z+N_Ch_b^2\Theta_Z+h_\tau^2\Theta_Z
\right)\tan\beta \, ,
\end{eqnarray}
\begin{eqnarray}
\beta_{\mu_1^2}^{(1)}&=&
-2\left[6\mu_1^2\lambda_1+\mu_2^2(2\lambda_3+\lambda_4)
+6m_{12}^2\lambda_6\right] -2\gamma_1^{(1)}\mu_1^2 \, ,
\\[2mm]
\beta_{\mu_2^2}^{(1)}&=&
-2\left[6\mu_2^2\lambda_2+\mu_1^2(2\lambda_3+\lambda_4)
+6m_{12}^2\lambda_7\right] -2\gamma_2^{(1)}\mu_2^2 \, ,
\\[2mm]
\beta_{m_{12}^2}^{(1)} &=&
2\left[-(\lambda_3+2\lambda_4+6\lambda_5)m^2_{12}
-3\lambda_6\mu_1^2-3\lambda_7\mu_2^2\right]
-\left(\gamma_1^{(1)}+\gamma_2^{(1)}\right)m^2_{12} \, . \nonumber \\
\end{eqnarray}

\vspace{-0.5cm}
\setcounter{equation}{0} 
\section{Two-Loop 2HDM RGEs}\label{RGE:2HDM2loop}

In  this appendix,  we present  the RGEs  of the  general 2HDM  at the
two-loop  order, as  derived  in \cite{Dev:2014yca,Lee:2015uza}.   For
definiteness, we  follow the conventions  of~\cite{Lee:2015uza}, where
$g_3\to   g_s$,    $g_2   \to   g$,   $g_1^2    \to   (5/3)g'^2   \,;$
$\lambda_{1,2,5}\to                              -\,2\lambda_{1,2,5}$,
$\lambda_{3,4,6,7}\to   -\lambda_{3,4,6,7}$.     The   two-loop   beta
functions for the quartic couplings are given by
\begin{eqnarray}
\beta_{\lambda_1}^{(2)}&=&
-\frac{291}{16}g^6+\frac{101}{16}g^4g'^2+\frac{191}{16}g^2g'^4
+\frac{131}{16}g'^6 
\nonumber \\ &&
+\frac{3}{16}g^4
\left[12h_b^2+4h_\tau^2-34\lambda_1+20(2\lambda_3+\lambda_4)\right]
-\frac{1}{8}g^2g'^2
\left[36h_b^2+44h_\tau^2-78\lambda_1-20\lambda_4\right]
\nonumber \\ &&
-\frac{1}{16}g'^4
\left[20h_b^2-100h_\tau^2-434\lambda_1-20(2\lambda_3+\lambda_4)\right]
\nonumber \\ &&
+8g_s^2h_b^2\left[4h_b^2+10\lambda_1\right]
-\frac{3}{4}g^2\left[-10\lambda_1(3h_b^2+h_\tau^2)
+4(36\lambda_1^2+4\lambda_3^2+4\lambda_3\lambda_4+\lambda_4^2+18\lambda_6^2)
\right]
\nonumber \\ &&
-\frac{1}{12}g'^2\left[h_b^2(16h_b^2-50\lambda_1)
+3h_\tau^2(-16h_\tau^2-50\lambda_1) \right.
\nonumber \\ && \hspace{1.3cm} \left.
+12(36\lambda_1^2+4\lambda_3^2+4\lambda_3\lambda_4+2\lambda_4^2
-4\lambda_5^2+18\lambda_6^2)\right]
\nonumber \\ &&
+6h_t^2\left[2\lambda_3^2+2\lambda_3\lambda_4+\lambda_4^2
+4\lambda_5^2+6\lambda_6^2 \right]
-\frac{3}{2}h_t^2h_b^2\left[4h_b^2+6\lambda_1\right]
\nonumber \\ &&
+\frac{3}{2}h_b^2\left[-20h_b^4-2h_b^2\lambda_1+96\lambda_1^2
+24\lambda_6^2 \right]
+\frac{1}{2}h_\tau^2\left[-20h_\tau^4-2h_\tau^2\lambda_1+96\lambda_1^2
+24\lambda_6^2 \right]
\nonumber \\ &&
-2\lambda_1\left[156\lambda_1^2+10\lambda_3^2+10\lambda_3\lambda_4
+6\lambda_4^2+28\lambda_5^2+159\lambda_6^2-3\lambda_7^2\right]
\nonumber \\ &&
-2\lambda_3\left[4\lambda_3^2+6\lambda_3\lambda_4
+ 8\lambda_4^2
+40\lambda_5^2+33\lambda_6^2+18\lambda_6\lambda_7+9\lambda_7^2\right]
\nonumber \\ &&
-2\lambda_4\left[3\lambda_4^2+44\lambda_5^2+35\lambda_6^2
+14\lambda_6\lambda_7+7\lambda_7^2\right]
-4\lambda_5\left[37\lambda_6^2+10\lambda_6\lambda_7+5\lambda_7^2\right]\;,
\end{eqnarray}
\begin{eqnarray}
\beta_{\lambda_2}^{(2)}&=&
-\frac{291}{16}g^6+\frac{101}{16}g^4g'^2+\frac{191}{16}g^2g'^4
+\frac{131}{16}g'^6
\nonumber \\ &&
+\frac{3}{16}g^4
\left[12h_t^2-34\lambda_2+20(2\lambda_3+\lambda_4)\right]
-\frac{1}{8}g^2g'^2
\left[84h_t^2-78\lambda_2-20\lambda_4\right]
\nonumber \\ &&
+\frac{1}{16}g'^4
\left[76h_t^2+434\lambda_2+20(2\lambda_3+\lambda_4)\right]
\nonumber \\ &&
+8g_s^2h_t^2\left[4h_t^2+10\lambda_2\right]
-\frac{3}{4}g^2\left[-30h_t^2\lambda_2
+4(36\lambda_2^2+4\lambda_3^2+4\lambda_3\lambda_4+\lambda_4^2+18\lambda_7^2)
\right]
\nonumber \\ &&
-\frac{1}{12}g'^2\left[-h_t^2(32h_t^2+170\lambda_2)
+12(36\lambda_2^2+4\lambda_3^2+4\lambda_3\lambda_4+2\lambda_4^2
-4\lambda_5^2+18\lambda_7^2)\right]
\nonumber \\ &&
-\frac{3}{2}h_t^2\left[20h_t^4+2h_t^2\lambda_2-96\lambda_2^2
-24\lambda_7^2 \right]
-\frac{3}{2}h_t^2h_b^2\left[4h_t^2+6\lambda_2\right]
\nonumber \\ &&
+(6h_b^2+2h_\tau^2) \left[2\lambda_3^2+2\lambda_3\lambda_4+\lambda_4^2
+4\lambda_5^2+6\lambda_7^2\right]
\nonumber \\ &&
-2\lambda_2\left[156\lambda_2^2+10\lambda_3^2+10\lambda_3\lambda_4
+6\lambda_4^2+28\lambda_5^2-3\lambda_6^2+159\lambda_7^2\right]
\nonumber \\ &&
-2\lambda_3\left[4\lambda_3^2+6\lambda_3\lambda_4
+8\lambda_4^2
+40\lambda_5^2+9\lambda_6^2+18\lambda_6\lambda_7+33\lambda_7^2\right]
\nonumber \\ &&
-2\lambda_4\left[3\lambda_4^2+44\lambda_5^2+7\lambda_6^2
+14\lambda_6\lambda_7+35\lambda_7^2\right]
-4\lambda_5\left[5\lambda_6^2+10\lambda_6\lambda_7+37\lambda_7^2\right]\;,
\end{eqnarray}
\begin{eqnarray}
\beta_{\lambda_3}^{(2)}&=&
-\frac{291}{8}g^6-\frac{11}{8}g^4g'^2-\frac{101}{8}g^2g'^4
+\frac{131}{8}g'^6
\nonumber \\ &&
+\frac{3}{4}g^4\left[3h_t^2+3h_b^2+h_\tau^2
+30(\lambda_1+\lambda_2)-\frac{37}{2}\lambda_3+10\lambda_4 \right]
\nonumber \\ &&
+\frac{1}{2}g^2g'^2\left[21h_t^2+9h_b^2+11h_\tau^2-10(\lambda_1+\lambda_2)
+\frac{11}{2}\lambda_3-6\lambda_4 \right]
\nonumber \\ &&
-\frac{1}{8}g'^4\left[-38h_t^2+10h_b^2-50h_\tau^2-60(\lambda_1+\lambda_2)
-197\lambda_3-20\lambda_4 \right]
\nonumber \\ &&
+8g_s^2
\left[8h_t^2h_b^2+5\lambda_3(h_t^2+h_b^2)\right]
\nonumber \\ &&
-6g^2\left[-\frac{5}{8}\lambda_3(3h_t^2+3h_b^2+h_\tau^2)
+6(\lambda_1+\lambda_2)(2\lambda_3+\lambda_4)+(\lambda_3-\lambda_4)^2
+18\lambda_6\lambda_7\right]
\nonumber \\ &&
-\frac{1}{3}g'^2\left[-4h_t^2h_b^2-\frac{5}{4}\lambda_3(17h_t^2+5h_b^2
+15h_\tau^2)\right.
\nonumber \\ && \hspace{1.3cm} \left.
+24(\lambda_1+\lambda_2)(3\lambda_3+\lambda_4)+6(\lambda_3^2-\lambda_4^2
+8\lambda_5^2+\lambda_6^2+16\lambda_6\lambda_7+\lambda_7^2) \right]
\nonumber \\ &&
+\frac{9}{2}\lambda_3\left[3h_t^4+3h_b^4+h_\tau^4\right]
-h_t^2h_b^2\left[36(h_t^2+h_b^2)-15\lambda_3\right]
\nonumber \\ &&
+6h_t^2\left[12\lambda_2\lambda_3+4\lambda_2\lambda_4+2\lambda_3^2
+\lambda_4^2+4\lambda_5^2+8\lambda_6\lambda_7+4\lambda_7^2\right]
\nonumber \\ &&
+(6h_b^2+2h_\tau^2)\left[12\lambda_1\lambda_3+4\lambda_1\lambda_4+2\lambda_3^2
+\lambda_4^2+4\lambda_5^2+4\lambda_6^2+8\lambda_6\lambda_7\right]
\nonumber \\ &&
-4(\lambda_1^2+\lambda_2^2)\left[15\lambda_3+4\lambda_4\right]
-4(\lambda_1+\lambda_2)\left[18\lambda_3^2+8\lambda_3\lambda_4
+7\lambda_4^2+36\lambda_5^2\right]
\nonumber \\ &&
-4\lambda_1\left[31\lambda_6^2+22\lambda_6\lambda_7+11\lambda_7^2\right]
-4\lambda_2\left[11\lambda_6^2+22\lambda_6\lambda_7+31\lambda_7^2\right]
\nonumber \\ &&
-\lambda_3\left[12\lambda_3^2+4\lambda_3\lambda_4+16\lambda_4^2+72\lambda_5^2
+60\lambda_6^2+176\lambda_6\lambda_7+60\lambda_7^2\right]
\nonumber \\ &&
-\lambda_4\left[12\lambda_4^2+176\lambda_5^2+68\lambda_6^2
+88\lambda_6\lambda_7+68\lambda_7^2\right]
\nonumber \\ &&
-2\lambda_5\left[68\lambda_6^2+72\lambda_6\lambda_7+68\lambda_7^2\right]\;,
\end{eqnarray}
\begin{eqnarray}
\beta_{\lambda_4}^{(2)}&=&
+14g^4g'^2+\frac{73}{2}g^2g'^4
+\lambda_4\left[-\frac{231}{8}g^4+\frac{157}{8}g'^4\right]
\nonumber \\ &&
-\frac{1}{2}g^2g'^2\left[42h_t^2+18h_b^2+22h_\tau^2-20\lambda_1
-20\lambda_2-4\lambda_3-\frac{51}{2}\lambda_4 \right]
\nonumber \\ &&
-8g_s^2\left[8h_t^2h_b^2-5\lambda_4(h_t^2+h_b^2)\right]
\nonumber \\ &&
-g^2\left[-\frac{15}{4}\lambda_4(3h_t^2+3h_b^2+h_\tau^2)
+18(2\lambda_3\lambda_4+\lambda_4^2+12\lambda_5^2+3\lambda_6^2+3\lambda_7^2)
\right]
\nonumber \\ &&
-\frac{1}{3}g'^2\left[4h_t^2h_b^2
-\frac{5}{4}\lambda_4(17h_t^2+5h_b^2+15h_\tau^2)
+12\lambda_4(2\lambda_1+2\lambda_2+\lambda_3+2\lambda_4)\right.
\nonumber \\ && \hspace{1.1cm} \left.
+6(32\lambda_5^2+7\lambda_6^2+4\lambda_6\lambda_7+7\lambda_7^2)
\right]
\nonumber \\ &&
-\frac{9}{2}\lambda_4\left[3h_t^4+3h_b^4+h_\tau^4\right]
+h_t^2h_b^2\left[24(h_t^2+h_b^2)-24\lambda_3-33\lambda_4\right]
\nonumber \\ &&
+12h_t^2\left[2\lambda_2\lambda_4+2\lambda_3\lambda_4+\lambda_4^2
+8\lambda_5^2+\lambda_6\lambda_7+5\lambda_7^2\right]
\nonumber \\ &&
+(12h_b^2+4h_\tau^2)\left[2\lambda_1\lambda_4+2\lambda_3\lambda_4
+\lambda_4^2+8\lambda_5^2+5\lambda_6^2+\lambda_6\lambda_7\right]
\nonumber \\ &&
-28\lambda_4\left[\lambda_1^2+\lambda_2^2\right]
-8(\lambda_1+\lambda_2)\left[10\lambda_3\lambda_4+5\lambda_4^2
+24\lambda_5^2\right]
\nonumber \\ &&
-4\lambda_1\left[37\lambda_6^2+10\lambda_6\lambda_7+5\lambda_7^2\right]
-4\lambda_2\left[5\lambda_6^2+10\lambda_6\lambda_7+37\lambda_7^2\right]
\nonumber \\ &&
-4\lambda_3\left[7\lambda_3\lambda_4+7\lambda_4^2+48\lambda_5^2
+18\lambda_6^2+20\lambda_6\lambda_7+18\lambda_7^2\right]
\nonumber \\ &&
-2\lambda_4\left[52\lambda_5^2
+34\lambda_6^2+80\lambda_6\lambda_7+34\lambda_7^2\right]
-32\lambda_5\left[5\lambda_6^2+6\lambda_6\lambda_7+5\lambda_7^2\right]\; ,
\end{eqnarray}
\begin{eqnarray}
\beta_{\lambda_5}^{(2)}&=&
+\lambda_5\left[-\frac{231}{8}g^4+\frac{19}{4}g^2g'^2+\frac{157}{8}g'^4\right]
+40g_s^2\lambda_5\left[h_t^2+h_b^2\right]
\nonumber \\ &&
-\frac{3}{8}g^2\left[-10\lambda_5(3h_t^2+3h_b^2+h_\tau^2)
+96\lambda_5(\lambda_3+2\lambda_4)+72(
\lambda_6^2 +\lambda_7^2)\right]
\nonumber \\ &&
-\frac{1}{24}g'^2\left[-10\lambda_5(17h_t^2+5h_b^2+15h_\tau^2)
-48\lambda_5(2\lambda_1+2\lambda_2-8\lambda_3-12\lambda_4) \right.
\nonumber \\ && \hspace{1.3cm} \left.
+48(5\lambda_6^2-\lambda_6\lambda_7+5\lambda_7^2) \right]
\nonumber \\ &&
-\frac{1}{2}\lambda_5\left[3h_t^4+3h_b^4+h_\tau^4\right]
-h_t^2\left[33h_b^2\lambda_5-12\lambda_5(2\lambda_2+2\lambda_3+3\lambda_4)
-6\lambda_7(\lambda_6+5\lambda_7)\right]
\nonumber \\ &&
+(6h_b^2+2h_\tau^2)\left[2\lambda_5(2\lambda_1+2\lambda_3+3\lambda_4)
+\lambda_6(5\lambda_6+\lambda_7)\right]
\nonumber \\ &&
-28\lambda_5\left[\lambda_1^2+\lambda_2^2\right]
-8\lambda_5(\lambda_1+\lambda_2)(10\lambda_3+11\lambda_4)
\nonumber \\ &&
-2\lambda_1\left[37\lambda_6^2+10\lambda_6\lambda_7+5\lambda_7^2\right]
-2\lambda_2\left[5\lambda_6^2+10\lambda_6\lambda_7+37\lambda_7^2\right]
\nonumber \\ &&
-2\lambda_3\left[14\lambda_3\lambda_5+38\lambda_4\lambda_5
+18\lambda_6^2+20\lambda_6\lambda_7+18\lambda_7^2\right]
\nonumber \\ &&
-2\lambda_4\left[16\lambda_4\lambda_5
+19\lambda_6^2+22\lambda_6\lambda_7+19\lambda_7^2\right]
+ 24\lambda_5^3
-8\lambda_5\left[9\lambda_6^2+21\lambda_6\lambda_7+9\lambda_7^2\right]\;,
\end{eqnarray}
\begin{eqnarray}
\beta_{\lambda_6}^{(2)}&=&
+\frac{1}{8}g^4\left[-141\lambda_6+90\lambda_7\right]
+\frac{1}{4}g^2g'^2\left[29\lambda_6+10\lambda_7\right]
+\frac{1}{8}g'^4\left[187\lambda_6+30\lambda_7\right]
+20g_s^2\lambda_6\left[h_t^2+3h_b^2\right]
\nonumber \\ &&
+\frac{9}{8}g^2\left[5\lambda_6(h_t^2+3h_b^2+h_\tau^2)
-96\lambda_6(\lambda_1+\lambda_5)
-16\lambda_6(\lambda_3+2\lambda_4)
-16\lambda_7(2\lambda_3+\lambda_4)\right]
\nonumber \\ &&
+\frac{1}{24}g'^2\left[5\lambda_6(17h_t^2+15h_b^2+45h_\tau^2)
-48\lambda_6(18\lambda_1+3\lambda_3+5\lambda_4+20\lambda_5) \right.
\nonumber \\ && \hspace{1.3cm} \left.
-48\lambda_7(6\lambda_3+4\lambda_4-2\lambda_5)
\right]
\nonumber \\ &&
-\frac{1}{4}\lambda_6\left[27h_t^4+84h_t^2h_b^2+33h_b^4+11h_\tau^4\right]
\nonumber \\ &&
+6h_t^2\left[\lambda_6(3\lambda_3+4\lambda_4+10\lambda_5)
+\lambda_7(6\lambda_3+4\lambda_4+ 4\lambda_5)\right]
\nonumber \\ &&
+(6h_b^2+2h_\tau^2)\lambda_6\left[24\lambda_1+3\lambda_3+4\lambda_4
+10\lambda_5 \right]
\nonumber \\ &&
-6\lambda_6\left[53\lambda_1^2-\lambda_2^2\right]
-4\lambda_1\lambda_6\left[33\lambda_3+35\lambda_4+74\lambda_5\right]
-4\lambda_2\lambda_6\left[9\lambda_3+7\lambda_4+10\lambda_5\right]
\nonumber \\ &&
-2\lambda_6\left[16\lambda_3^2+34\lambda_3\lambda_4+72\lambda_3\lambda_5
+17\lambda_4^2+76\lambda_4\lambda_5+72\lambda_5^2\right]
\nonumber \\ &&
-4(\lambda_1+\lambda_2)\lambda_7\left[9\lambda_3+7\lambda_4+10\lambda_5\right]
\nonumber \\ &&
-2\lambda_7\left[18\lambda_3^2+28\lambda_3\lambda_4+40\lambda_3\lambda_5
+17\lambda_4^2+44\lambda_4\lambda_5+84\lambda_5^2\right]
\nonumber \\ &&
-3\left[37\lambda_6^3+42\lambda_6^2\lambda_7
+11\lambda_6\lambda_7^2+14\lambda_7^3\right]\;,
\end{eqnarray}
\begin{eqnarray}
\beta_{\lambda_7}^{(2)}&=&
+\frac{1}{8}g^4\left[90\lambda_6-141\lambda_7\right]
+\frac{1}{4}g^2g'^2\left[10\lambda_6+29\lambda_7\right]
+\frac{1}{8}g'^4\left[30\lambda_6+187\lambda_7\right]
+20g_s^2\lambda_6\left[3h_t^2+h_b^2\right]
\nonumber \\ &&
+\frac{3}{8}g^2\left[5\lambda_7(9h_t^2+3h_b^2+h_\tau^2)
-288\lambda_7(\lambda_2+\lambda_5)
-48\lambda_3(2\lambda_6+\lambda_7)
-48\lambda_4(\lambda_6+2\lambda_7)\right]
\nonumber \\ &&
+\frac{1}{24}g'^2\left[5\lambda_7(51h_t^2+5h_b^2+15h_\tau^2)
-48\lambda_6(6\lambda_3+4\lambda_4-2\lambda_5) \right.
\nonumber \\ && \hspace{1.3cm} \left.
-48\lambda_7(18\lambda_2+3\lambda_3+5\lambda_4+20\lambda_5)\right]
\nonumber \\ &&
-\frac{3}{4}\lambda_7\left[11h_t^4+28h_t^2h_b^2+9h_b^4+3h_\tau^4\right]
\nonumber \\ &&
+6h_t^2\lambda_7\left[24\lambda_2
+3\lambda_3+4\lambda_4+10\lambda_5\right]
\nonumber \\ &&
+(6h_b^2+2h_\tau^2)\left[
\lambda_6(6\lambda_3+4\lambda_4 + 4\lambda_5)
+\lambda_7(3\lambda_3+4\lambda_4 +10\lambda_5) \right]
\nonumber \\ &&
-6\lambda_7\left[-\lambda_1^2+53\lambda_2^2\right]
-4\lambda_1\lambda_7\left[9\lambda_3+7\lambda_4+10\lambda_5\right]
-4\lambda_2 \lambda_7
\left[33\lambda_3+35\lambda_4+74\lambda_5\right]
\nonumber \\ &&
-2\lambda_7\left[16\lambda_3^2+34\lambda_3\lambda_4+72\lambda_3\lambda_5
+17\lambda_4^2+76\lambda_4\lambda_5+72\lambda_5^2\right]
\nonumber \\ &&
-4(\lambda_1+\lambda_2)
\lambda_6 \left[9\lambda_3+7\lambda_4+10\lambda_5\right]
\nonumber \\ &&
-2\lambda_6\left[18\lambda_3^2+28\lambda_3\lambda_4+40\lambda_3\lambda_5
+17\lambda_4^2+44\lambda_4\lambda_5+84\lambda_5^2\right]
\nonumber \\ &&
-3\left[14\lambda_6^3+11\lambda_6^2\lambda_7
+42\lambda_6\lambda_7^2+37\lambda_7^3\right]\; .
\end{eqnarray}

In addition, the two-loop beta functions for the gauge and  the
supersymmetric Yukawa couplings may be listed as follows:
\begin{eqnarray}
\beta^{(2)}_{g'}=\frac{5}{3}g'^3\Bigg(
\frac{44}{5}g_s^2+\frac{18}{5}g^2+\frac{104}{15}g'^2
-\frac{17}{10}h_t^2-\frac{1}{2}h_b^2-\frac{3}{2}h_\tau^2
\Bigg)\,,
\end{eqnarray}
\begin{eqnarray}
\beta^{(2)}_{g}=g^3\Bigg(
12g_s^2+8g^2+2g'^2
-\frac{3}{2}h_t^2-\frac{3}{2}h_b^2-\frac{1}{2}h_\tau^2
\Bigg)\,,
\end{eqnarray}
\begin{eqnarray}
\beta^{(2)}_{g_s}=g_s^3\Bigg(
-26g_s^2+\frac{9}{2}g^2+\frac{11}{6}g'^2
-2h_t^2-2h_b^2
\Bigg)\,,
\end{eqnarray}
\begin{eqnarray}
\beta^{(2)}_{h_t}&=&h_t\Bigg[
-108g_s^4+9g_s^2g^2+\frac{19}{9}g_s^2g'^2
-\frac{21}{4}g^4-\frac{3}{4}g^2g'^2+\frac{1267}{216}g'^4
\nonumber \\ &&
+g_s^2\left(36h_t^2+\frac{16}{3}h_b^2\right)
+\frac{3}{16}g^2\left(75h_t^2+11h_b^2\right)
+\frac{1}{48}g'^2\left(393h_t^2-\frac{41}{3}h_b^2\right)
\nonumber \\ &&
-12h_t^4-\frac{5}{2}h_t^2h_b^2-\frac{5}{2}h_b^4
-\frac{3}{4}h_b^2h_\tau^2
+12h_t^2\lambda_2+2h_b^2(\lambda_3-\lambda_4)
\nonumber \\ &&
+6\lambda_2^2+\lambda_3^2+\lambda_3\lambda_4+\lambda_4^2
+6\lambda_5^2+\frac{3}{2}\lambda_6^2+\frac{9}{2}\lambda_7^2
\Bigg]\,,
\end{eqnarray}
\begin{eqnarray}
\beta^{(2)}_{h_b}&=&h_b\Bigg[
-108g_s^4+9g_s^2g^2+\frac{31}{9}g_s^2g'^2
-\frac{21}{4}g^4-\frac{9}{4}g^2g'^2+\frac{113}{216}g'^4
\nonumber \\ &&
+g_s^2\left(\frac{16}{3}h_t^2+36h_b^2\right)
+\frac{3}{16}g^2\left(11h_t^2+75h_b^2+10h_\tau^2\right)
-\frac{1}{144}g'^2\left(53h_t^2-711h_b^2-450h_\tau^2\right)
\nonumber \\ &&
-\frac{5}{2}h_t^4-\frac{5}{2}h_t^2h_b^2-12h_b^4
-\frac{9}{4}h_b^2h_\tau^2
-\frac{9}{4}h_\tau^4
+12h_b^2\lambda_1
+2h_t^2(\lambda_3-\lambda_4)
\nonumber \\ &&
+6\lambda_1^2+\lambda_3^2+\lambda_3\lambda_4+\lambda_4^2
+6\lambda_5^2+\frac{9}{2}\lambda_6^2+\frac{3}{2}\lambda_7^2
\Bigg]\,,
\end{eqnarray}
\begin{eqnarray}
\beta^{(2)}_{h_\tau}&=&h_\tau\Bigg[
20g_s^2h_b^2
-\frac{21}{4}g^4+\frac{9}{4}g^2g'^2+\frac{161}{8}g'^4
\nonumber \\ &&
+\frac{15}{16}g^2\left(6h_b^2+11h_\tau^2\right)
+\frac{1}{48}g'^2\left(50h_b^2+537h_\tau^2\right)
\nonumber \\ &&
-\frac{9}{4}h_t^2h_b^2
-\frac{27}{4}h_b^4
-\frac{27}{4}h_b^2h_\tau^2
-3h_\tau^4
+12h_\tau^2\lambda_1
\nonumber \\ &&
+6\lambda_1^2+\lambda_3^2+\lambda_3\lambda_4+\lambda_4^2
+6\lambda_5^2+\frac{9}{2}\lambda_6^2+\frac{3}{2}\lambda_7^2
\Bigg]\,.
\end{eqnarray}

Finally, the two-loop anomalous dimensions for the Higgs doublets are given by
\begin{eqnarray}
\gamma^{(2)}_1&=&
\frac{435}{32}g^4-\frac{3}{16}g^2g'^2-\frac{149}{32}g'^4-20g_s^2h_b^2
-\frac{15}{8}g^2\left(3h_b^2+h_\tau^2\right)
-\frac{25}{24}g'^2\left(h_b^2+3h_\tau^2\right)
\nonumber \\ &&
+\frac{9}{4}h_t^2h_b^2+\frac{27}{4}h_b^4+\frac{9}{4}h_\tau^4
-6\lambda_1^2-\lambda_3^2-\lambda_3\lambda_4-\lambda_4^2
-6\lambda_5^2-\frac{9}{2}\lambda_6^2-\frac{3}{2}\lambda_7^2
\nonumber \\ &&
-\frac{3}{2}t_\beta\left[2\lambda_1\lambda_6+2\lambda_2\lambda_7
+(\lambda_3+\lambda_4+2\lambda_5)(\lambda_6+\lambda_7)\right]\,,
\end{eqnarray}
\begin{eqnarray}
\gamma^{(2)}_2&=&
\frac{435}{32}g^4-\frac{3}{16}g^2g'^2-\frac{149}{32}g'^4
-h_t^2\left(20g_s^2+\frac{45}{8}g^2+\frac{85}{24}g'^2\right)
\nonumber \\ &&
+\frac{27}{4}h_t^4+\frac{9}{4}h_b^2h_t^2
-6\lambda_2^2-\lambda_3^2-\lambda_3\lambda_4-\lambda_4^2
-6\lambda_5^2-\frac{3}{2}\lambda_6^2-\frac{9}{2}\lambda_7^2
\nonumber \\ &&
-\frac{3}{2}t_\beta^{-1}\left[2\lambda_1\lambda_6+2\lambda_2\lambda_7
+(\lambda_3+\lambda_4+2\lambda_5)(\lambda_6+\lambda_7)\right]\,.
\end{eqnarray}

\setcounter{equation}{0}
\section{Threshold corrections to $\lambda_i$ at $M_S$} 

At the  soft SUSY-breaking scale  $Q = M_S$,  we need to  consider the
threshold  corrections to  quartic couplings  due to  third-generation
sfermions. These  are derived  in~\cite{Lee:2015uza}, which  we extend
here to include CP-violating phases.

The quartic couplings $\lambda_{i}$ with $i=1-7$ at the RG scale $Q = M_S$ are
given by
\begin{equation}
\lambda_i(M_S) = \lambda_i^{(0)}+\sum_{n=1,2} \kappa^n
\Delta^{(n)}\lambda_i\,,
\end{equation}
where
\begin{eqnarray}
&&
\lambda_1^{(0)}\ =\ \lambda_2^{(0)}\ =\ -\, \frac{1}{8}\, (g^2 + g'^2)\,, \ \
\lambda_3^{(0)} = -\frac{1}{4}\, (g^2 -g'^2)\,, \  \
\lambda_4^{(0)}\ =\ \frac{1}{2}\, g^2\,, \nonumber \\[2mm]
&&
\lambda_5^{(0)}\ =\ \lambda_6^{(0)}\ =\ \lambda_7^{(0)}\ =\ 0\, ,
\end{eqnarray}
and the one- and two-loop threshold corrections are
\footnote{Here all the mass parameters are dimensionsless 
and normalized to the SUSY scale $M_S$:
$\widehat \mu=\mu/M_S$,
$\widehat A_{t,b,\tau}=A_{t,b,\tau}/M_S$, and
$\widehat M_3=M_3/M_S$.}
\begin{eqnarray}
\Delta^{(1)}\lambda_1 &=&
 \frac{1}{4}|h_t|^4 |\widehat\mu|^4
-3 |h_b|^4 |\widehat A_b|^2\left(1-\frac{|\widehat A_b|^2}{12}\right)
- |h_\tau|^4 |\widehat A_\tau|^2
\left(1-\frac{|\widehat A_\tau|^2}{12}\right) \nonumber \\
&&
-\frac{g^2+g'^2}{8}\left(
3|h_t|^2 |\widehat\mu|^2-3|h_b|^2 |\widehat A_b|^2
-|h_\tau|^2 |\widehat A_\tau|^2\right) \nonumber \\
&&
+\frac{g^2+g'^2}{24}\left(
3|h_t|^2 |\widehat\mu|^2+3|h_b|^2 |\widehat A_b|^2
+|h_\tau|^2 |\widehat A_\tau|^2\right) \,,  \\[2mm]
\Delta^{(1)}\lambda_2 &=&
-3 |h_t|^4 |\widehat A_t|^2\left(1-\frac{|\widehat A_t|^2}{12}\right)
+\frac{1}{4}|h_b|^4 |\widehat\mu|^4
+\frac{1}{12}|h_\tau|^4 |\widehat\mu|^4 \nonumber \\
&&
+\frac{g^2+g'^2}{8}\left(
3|h_t|^2 |\widehat A_t|^2-3|h_b|^2 |\widehat \mu|^2
-|h_\tau|^2 |\widehat \mu|^2\right) \nonumber \\
&&
+\frac{g^2+g'^2}{24}\left(
3|h_t|^2 |\widehat A_t|^2+3|h_b|^2 |\widehat \mu|^2
+|h_\tau|^2 |\widehat \mu|^2\right)  \,, \\[2mm]
\Delta^{(1)}\lambda_3 &=&
-\frac{1}{6} |\mu|^2 \left[3|h_t|^4(3-|\widehat A_t|^2)
+3|h_b|^4(3-|\widehat A_b|^2)
+|h_\tau|^4(3-|\widehat A_\tau|^2)\right] \nonumber \\
&&
-\frac{1}{2}|h_t|^2|h_b|^2\left[ 
3|\widehat A_t+\widehat A_b|^2
-||\widehat \mu|^2-\widehat A_t \widehat A_b^*|^2
-6|\widehat \mu|^2\right]  \nonumber \\
&&
+\frac{g^2-g'^2}{8}\left[
 3|h_t|^2 (|\widehat A_t|^2-|\widehat \mu|^2)
+3|h_b|^2 (|\widehat A_b|^2-|\widehat \mu|^2)
+|h_\tau|^2 (|\widehat A_\tau|^2-|\widehat \mu|^2)\right] \nonumber \\
&&
+\frac{g^2-g'^2}{24}\left[
 3|h_t|^2 (|\widehat A_t|^2+|\widehat \mu|^2)
+3|h_b|^2 (|\widehat A_b|^2+|\widehat \mu|^2)
+|h_\tau|^2 (|\widehat A_\tau|^2+|\widehat \mu|^2)\right]  \,, 
\nonumber \\ \\[2mm]
\Delta^{(1)}\lambda_4 &=& 
-\frac{1}{6} |\mu|^2 \left[3|h_t|^4(3-|\widehat A_t|^2)
+3|h_b|^4(3-|\widehat A_b|^2)
+|h_\tau|^4(3-|\widehat A_\tau|^2)\right] \nonumber \\
&&
+\frac{1}{2}|h_t|^2|h_b|^2\left[
3|\widehat A_t+\widehat A_b|^2
-||\widehat \mu|^2-\widehat A_t \widehat A_b^*|^2
-6|\widehat \mu|^2\right]  \nonumber \\
&&
-\frac{g^2}{4}\left[
 3|h_t|^2 (|\widehat A_t|^2-|\widehat \mu|^2)
+3|h_b|^2 (|\widehat A_b|^2-|\widehat \mu|^2)
+|h_\tau|^2 (|\widehat A_\tau|^2-|\widehat \mu|^2)\right] \nonumber \\
&&
-\frac{g^2}{12}\left[
 3|h_t|^2 (|\widehat A_t|^2+|\widehat \mu|^2)
+3|h_b|^2 (|\widehat A_b|^2+|\widehat \mu|^2)
+|h_\tau|^2 (|\widehat A_\tau|^2+|\widehat \mu|^2)\right]  \,, \\[2mm]
\Delta^{(1)}\lambda_5 &=& 
\frac{1}{12}\left[
 3h_t^4 \widehat \mu^2 \widehat A_t^2
+3h_b^4 \widehat \mu^2 \widehat A_b^2
+h_\tau^4 \widehat \mu^2 \widehat A_\tau^2 \right]  \,, \\[2mm]
\Delta^{(1)}\lambda_6 &=& 
-\frac{1}{6}\left[
 3h_t^4 |\widehat \mu|^2 \widehat \mu \widehat A_t
+3h_b^4 \widehat \mu \widehat A_b (|\widehat A_b|^2-6)
+h_\tau^4 \widehat \mu \widehat A_\tau (|\widehat A_\tau|^2-6) \right]  \,, 
\\[2mm]
\Delta^{(1)}\lambda_7 &=&
-\frac{1}{6}\left[
 3h_t^4 \widehat \mu \widehat A_t (|\widehat A_t|^2-6)
+3h_b^4 |\widehat \mu|^2 \widehat \mu \widehat A_b
+h_\tau^4 |\widehat \mu|^2 \widehat \mu \widehat A_\tau\right]  \,,
\end{eqnarray}
where $h_{t,b,\tau}=h_{t,b,\tau}^{\rm MSSM}$ at the RG scale $Q=M_S$.

The two-loop corrections of ${\cal O}(|h_t|^4g_s^2)$ are given by
\begin{eqnarray}
\Delta^{(2)} \lambda_1 &=&
 \frac{2}{3} |h_t|^4g_s^2 |\widehat\mu|^4\,, \\[2mm]
\Delta^{(2)} \lambda_2 &=&
-8|h_t|^4g_s^2 \left[
-2\real(\widehat A_t \widehat M_3^*) 
+\frac{1}{3} |\widehat A_t|^2 \real(\widehat A_t \widehat M_3^*)
-\frac{1}{12} |\widehat A_t|^4 \right]\,, ~~~~~~~~~~~~~~ \\[2mm]
\Delta^{(2)} \lambda_3 &=& \Delta^{(2)} \lambda_4 =
-\frac{8}{3} |h_t|^4g_s^2 |\widehat\mu|^2\left[
\real(\widehat A_t \widehat M_3^*) 
-\frac{1}{2} |\widehat A_t|^2 \right] \,, \\[2mm]
\Delta^{(2)} \lambda_5 &=& 
-\frac{4}{3} |h_t|^4g_s^2 \widehat\mu \widehat A_t \left[
\widehat\mu \widehat M_3
-\frac{1}{2} \widehat\mu \widehat A_t \right] \,, \\[2mm]
\Delta^{(2)} \lambda_6 &=& 
-\frac{4}{3} |h_t|^4g_s^2 |\widehat\mu|^2  \left[
-\widehat\mu \widehat M_3
+\widehat\mu \widehat A_t \right] \,, \\[2mm]
\Delta^{(2)} \lambda_7 &=& 
-4 |h_t|^4g_s^2 \left[
 2\widehat\mu \widehat M_3 
-\frac{1}{3}\widehat\mu \widehat A_t^2 \widehat M_3^*
-\frac{2}{3}|\widehat A_t|^2 \widehat\mu \widehat M_3
+\frac{1}{3}|\widehat A_t|^2 \widehat\mu \widehat A_t \right] \,. 
\end{eqnarray}


\end{appendix}

\medskip

\end{document}